\documentclass[aps,prb,showpacs,twocolumn,superscript,superscriptaddress]{revtex4-1}
\usepackage{bm,color,amsmath,amssymb,mathrsfs,latexsym,graphicx,psfrag,mathtools}
\usepackage{graphicx}
\usepackage{bm}
\usepackage{float}
\usepackage{multirow}
\usepackage[T1]{fontenc}
\usepackage{mathrsfs}
\usepackage[colorlinks=true,
            linkcolor = blue,
            urlcolor  = blue,
            citecolor = blue,
            anchorcolor = blue]{hyperref}
\usepackage{url}
\usepackage[utf8]{inputenc}
\usepackage{xcolor}
\usepackage{bm,upgreek,bm,appendix,soul,color,graphicx}
\usepackage[normalem]{ulem}

\usepackage{xcolor}
\usepackage{amsmath}
\usepackage{epsfig}
\usepackage{epstopdf}
\usepackage{comment}
\usepackage{ulem}

\newcommand{\bea}{\begin{eqnarray}}
\newcommand{\eea}{\end{eqnarray}}

\begin{document}

\bibliographystyle{unsrt}

\title{Time Reversal Symmetry Breaking and {\it Fragile Magnetic Superconductors}}
\author{Warren E. Pickett }
\affiliation{Department of Physics and Astronomy \\
University of California Davis\\
Davis CA 95616}

\begin{abstract}
Roughly twenty reports (as of 2026) of time-reversal-symmetry breaking (TRSB) 
states in low critical temperature (T$_c$) superconducting (SC), 
otherwise conventional Fermi liquid, metals have emerged primarily 
from muon spin relaxation ($\mu$SR) data but a few reports from polar Kerr
effect and magnetization. For each probe, the inferred fields are
not far above the lower limit of detection, corresponding to  
magnetizations (if uniform) of no more than 10$^{-3}$ $\mu_B$/atom. 
These materials, with ``TRSB'' interpretation due to a spontaneous magnetic
polarization below T$_c$, thus comprise a new class of 
{\it fragile magnetic superconductors}, modeled as triplet pairing.
The measured SC state properties, excepting only increased depolarization below
T$_c$, are representative of low T$_c$ singlet BCS SCs, not showing 
unusual coherence lengths or critical fields or the sensitivity to
defects characteristic of triplet states.
While it is recognized that the muon does affect the sample by displacing
nearby atoms and impacting magnetic interaction parameters, 
the measurement process, changing the system from
sample $\rightarrow$ sample+$\mu^+$ thereby breaking TRS, 
may deserve further scrutiny; specifically, the muon field induced supercurrent will produce a
magnetic field in the region of the muon. 
An overview of the environment of the muon,
from the normal state to the superfluid state is given, 
with attention to the muon-induced supercurrent and 
Yu-Shiba-Rusinov gap states, both of which provide coupling 
of the muon moment to the superconducting gap. Variation of reports
across sample quality is also referenced and discussed.
The unusual topological superconductor LaNiGa$_2$, currently modeled as
non-unitary triplet, is used as a case study.
Supposing that the prevailing $\mu$SR inference 
of a small spontaneous field within the bulk of the
SC obtains (recent NMR data provides a different picture), 
the current theoretical picture of triplet pairing is discussed
and an alternative singlet scenario for LaNiGa$_2$, based on 
symmetry breaking of an inversion-related pair
of Fermi surface Dirac points arising from its non-symmorphic symmetry, is outlined.
\end{abstract}
\date{\today}
\maketitle

\tableofcontents{}
\newpage

\section{Introduction}
\label{sec:intro}
A recent report on {\it nanostructured superconductivity}\cite{Lang2023}
giving an overview of the various ways in which the superconducting state
can be distorted when the bulk condensate is impeded by structural 
disruptions, begins with the sentence
``the relevant length scales for superconductivity (SC) are of the order of 
nanometers." This statement is appropriate for the emphasis of
that paper. Much of the theoretical description and
experimental interpretation of superconducting (SC) properties 
is done at the Ginzburg-Landau level 
of the coherence length (several to hundreds of nanometers) and the penetration 
depth (typically hundreds of nanometers or more), or incorporating the superfluid density 
at the London theory level. 

An important development in SC theory (BCS-Migdal-Eliashberg, and even beyond) is that
the materials-level theory of the singlet SC gap 2$\Delta$, 
the critical temperature $T_c$, the coherence length $\xi$ and
penetration depth $\lambda$ (both at zero temperature unless otherwise noted),
due to phonon exchange is formulated at the atomic 
level and is accurate to any realistic expectation.\cite{Pickett2023,isome}
The study of unconventional SC order parameters has progressed considerably 
on the basis of symmetry requirements satisfying both theoretical constraints and 
experimental data, with some of them assigned on the basis of muon spin
relaxation spectroscopy. There are largely unstudied sub-nm electronic processes 
near the muon in these metals that may benefit understanding of the muon
region as well as adding 
clarification for exotic pairing and condensate formation. 

Over the past three decades around twenty low-T$_c$ SCs, whose normal 
state properties and SC parameters are successfully understood
on the basis of weak-coupling Fermi liquids with singlet pairing, have been
characterized as time-reversal symmetry breaking (TRSB)  due to 
enhanced muon spin polarization interpreted as due to an emergent magnetic field, 
 as magnetization is the most common manifestation (but not the 
only) of TRSB.
Muon spin relaxation ($\mu$SR) determination of enhanced depolarization
below T$_c$,  a few polar Kerr effect (PKE) measurements and one 
magnetization study have also detected signatures of SC-induced changes
at, or near, their low T$_c$, almost all with T$_c$ of 2K or less.  

TRSB order parameters require exotic triplet pairing, which has attracted
considerable theoretical attention.\cite{Mackenzie2003,Stewart1984,
Schenck1985,Rudd1998,Annett1990,Sigrist1991,Annett1995,Tsuneto1998,Mineev1999,
Sigrist2005,Sigrist2009,Mackenzie2003,Wysokinski2019,Ramires2022,
Blount1985} The study of triplet superconductivity flowered
after the discovery of unusual SC in heavy fermion SCs, which
is not the subject of this paper. 
Similar effects have been detected in correlated electron 
superconductors, where unconventional order parameters are not
so unexpected. In both class of SCs sample differences have been reported.
 
The focus in this paper will be on exploring processes relating 
to the depolarization of the muon spin arising below T$_c$.
There have  been continuous extensions of the interpretation of
the $\mu$SR data, both in understanding the local perturbation
due to the implanted muon, and the treatment of thermal activity
of nearby magnetic moments. After recounting the progression 
of theory is probing the region around a muon at an interstitial 
site, this manuscript
addresses some near-muon processes that remain to be understood.  
The example of LaNiGa$_2$ will be given attention in this article
as a case study.\cite{Hillier2012,Weng2016,Ghosh2020,Grin1982,
Staab2022,HLi2025,Badger2022,Quan2022,Sherpa2023,Sundar2024,Ghimire2024}

There are a number of other relevant references.
A pedagogical, handbook-style monograph on muon spin rotation
spectroscopy in solids was published
by Schenck\cite{Schenck1985} in 1985, 
followed by a number of books, lecture notes,
and reports on $\mu$SR spectroscopy.\cite{Slichter1980,Brewer1981,
Brewer1993,Lee1999,Blundell1999,Yaouanc2011,Blundell2021}
The $\mu$SR experiment and analysis has been described by 
Blundell,\cite{Blundell1999}
a  view of $\mu$SR theory and data on selected
materials by Hillier {\it et al.},\cite{Hillier2022},
and a broad discussion provided in a monograph by
Blundell and co-authors.\cite{Blundell2021}
For recent techniques, Blundell and Lancaster provided\cite{Blundell2023} a 
description of a `DFT+$\mu$' method (density functional theory treatment
including the interstitial muon) for
finding the muon stopping position(s) in a crystal, 
with a guide on interface `MuFinder' provided by Huddart {\it et al.}~\cite{Huddart2022}
Calculation of the muon anharmonicity and zero point 
quantum positional uncertainty (QPU)
has been reported for several 
materials.\cite{Gomilsek2023,Errea2013,Monacelli2021,Errea2020,Errea2016,Onuorah2019} 

After discovery of heavy fermion SCs, Heffner reviewed 
in 1992 $\mu$SR studies of this class of quantum materials,
with emphasis on uranium superconductors.~\cite{Heffner1992}
This class of highly unconventional materials, discovered in the 
early 1980s, have strongly renormalized properties in the normal
state and are not the topic of this article. 
An extensive review of defect-induced properties in such materials, and
in conventional superconductors, was provided by Balatsky,
Vekhter, and Zhu in 2006.\cite{Balatsky2006}

Reviews on exotic order parameters more generally are relevant to 
discussions in this paper. Earlier work on isotropic superfluid $^3$He 
required adaptation to crystal systems. In 1991 Sigrist and Ueda\cite{Sigrist1991} 
provided a review and extension of the theory of unconventional SC states, 
extending from the generalization of BCS theory\cite{BCS1957} to symmetry 
classification and its relation to Ginzburg-Landau theory,\cite{Ginzburg1950} 
to symmetry-breaking including the non-unitary possibility for triplet 
superconductors, to crystal symmetry lowering effects (structural transformations, 
consequences of surfaces and interfaces, and more). Sigrist has provided 
following reviews on broken time-reversal symmetry,\cite{Sigrist2000}, on 
unconventional SCs,\cite{Sigrist2005} and on an extension specifically aimed 
at non-centrosymmetric SCs.\cite{Sigrist2009} Wysokinski provided in 2019 an 
overview\cite{Wysokinski2019} of time-reversal symmetry breaking, 
with focus on Sr$_2$RuO$_4$.

An overview of the interplay between inhomogeneities and SC order parameters,
especially TRSB ones, by Andersen, Kreisel, and Hirschfeld, addressed
topics of relevance to $\mu$SR experiments\cite{Andersen2023} and relevant 
to the description in this article. They provide a brief but 
informative description of the zero field $\mu$SR experiment that has
provided evidence of TRSB SC phases in several otherwise conventional
intermetallic compounds.  Among the situations they discuss is that defects
in a TRSB SC can produce local magnetic fields from spin disruption and
from orbital currents. The current article will provide discussion of effects
of magnetic impurities, viz. the muon, in a generic Fermi liquid superconductor, leaving interaction of impurities with exotic OPs to specific treatments of such cases, cited above.

The organization of the paper is this. Background information on the 
muon's magnetic moment, the resulting field and induced magnetization,
and necessary quantum effects and implications of symmetry, for the
normal state, are presented in Sec.~\ref{sec:near-field}. Due to the 
breadth of the topics and the number of questions raised, this 
manuscript is more a set of accumulated notes and less of a
research paper ({\it i.e.} Introduction, Methods, Results, Summary.)
Near-muon quantum effects -- positional uncertainty, non-linear
susceptibility, electron pair correlation -- in Sec.~\ref{sec:QuantumEffects}
are addressed and suggested to regularize some difficulties of the semi-quantum
formalism. 
This beginning is followed by a listing of 19 reported TRSB materials 
with weak coupling and Fermi
liquid normal states, provided in Sec.~\ref{sec:TRSB}.

Passing into the SC state is the topic of Sec.~\ref{sec:superstate}, 
with the muon-induced supercurrent - magnetic field connections providing a
complex picture of the SC near-field. 
Effects of Kondo and related YSR states (magnetic states bound
within the SC gap) are included in this section. 
Issues concerning pairing
symmetry -- triplet versus singlet pairing, or other possible forms
of symmetry breaking --form the focus of Sec.~\ref{sec:OP}. 
Comments on the effects of sample
variability on identification of TRSB are included in 
Sec.~\ref{sec:sample}, after which the `case study' of LaNiGa$_2$
is presented in Sec.~\ref{sec:casestudy}. Sec.~\ref{sec:summary}
provides a Summary of the main points of the paper.

A few relevant discussions or numerical data
are relegated to appendices: the historical development of the
theory of supercurrent, Kondo impurity effects, material parameters
of LaNiGa$_2$, triplet order parameters. The Supplementary Material (SM) provides
the full microscopic non-relativistic Hamiltonian of
the muon+sample laid out with some discussion, but
manipulations of this Hamiltonian in this paper are few. Because supercurrents
are stimulated by vector potentials, the Hamiltonian is provided at that level.
The setup of the $\mu$SR experiment is given a non-expert
overview, including some discussion of the analysis of
data; the full analysis is no doubt more involved. 
Considerations of possible superconducting order parameters form the discussions
in Appendix \ref{app:tripletOP}, with application
to LaNiGa$_2$ in Sec.~\ref{subsec:scenario}: (possible) singlet versus triplet,
unitary versus non-unitarity, and the currently proposed
model of triplet pairing for two materials, and
the special role that LaNiGa$_2$ might play.


\section{Muon disturbances: electronic, magnetic}
\label{sec:near-field}
Upon implantation, the muon impacts the near-field region in a few ways.
Examination of the full nuclear+electronic+muon microscopic Hamiltonian 
of the coupled system in the SM Sec:~\ref{SM:hamiltonian} may be useful. 
A key topic will be the
effect, and the response, of the muon moment $\vec\mu$ and its vector
potential with resulting dipolar magnetic field\cite{Sautbekov2019}
 $\vec B^{\mu}$, with its charge, site symmetry, and environment
 playing parts. Some useful points are addressed more fully in the
Supplemental Material (SM). An amusing aside is the characterization 
of the origin of point particle moments by J. D. Jackson\cite{Jackson1977}:\\
\hskip 2mm
{\it ``Usually the books are a little vague about the nature of these
intrinsic magnetic moments, letting the word `intrinsic' imply 
that it is beyond the realms
of present knowledge or none of your business, or both.''}\\
He went on in his CERN document to describe the theoretical
description of intrinsic magnetic dipole moments, 
and that point (particle) dipoles must be regarded
as the limit of a current loop with area going to zero rather
than as north and south monopoles coalescing, 
consistent with experimental data from elementary particle studies. 

Throughout this paper $cgs-gaussian$
units will be used, for which the magnetic susceptibility $\chi$ is unitless.
Also, in non-magnetic metals the Pauli paramagnetic $\chi$ is of 
the order of 10${^-3}$  or less, so
$\vec B$ and the field intensity $\vec H$ ($\vec B=\vec H+4\pi\vec M=
(1+4\pi\chi)\vec H$) are effectively interchangeable except near the muon.
The electron mass will be denoted by $m$, the muon mass by $m_{\mu}$, 
and $m^*$ an electron effective mass. Electronic magnetization will be
denoted by $\vec m(\vec r)$, which by the context should not be
confused as a mass. The near-field involves several aspects.

\begin{figure}
 \includegraphics[width=1.3\columnwidth]{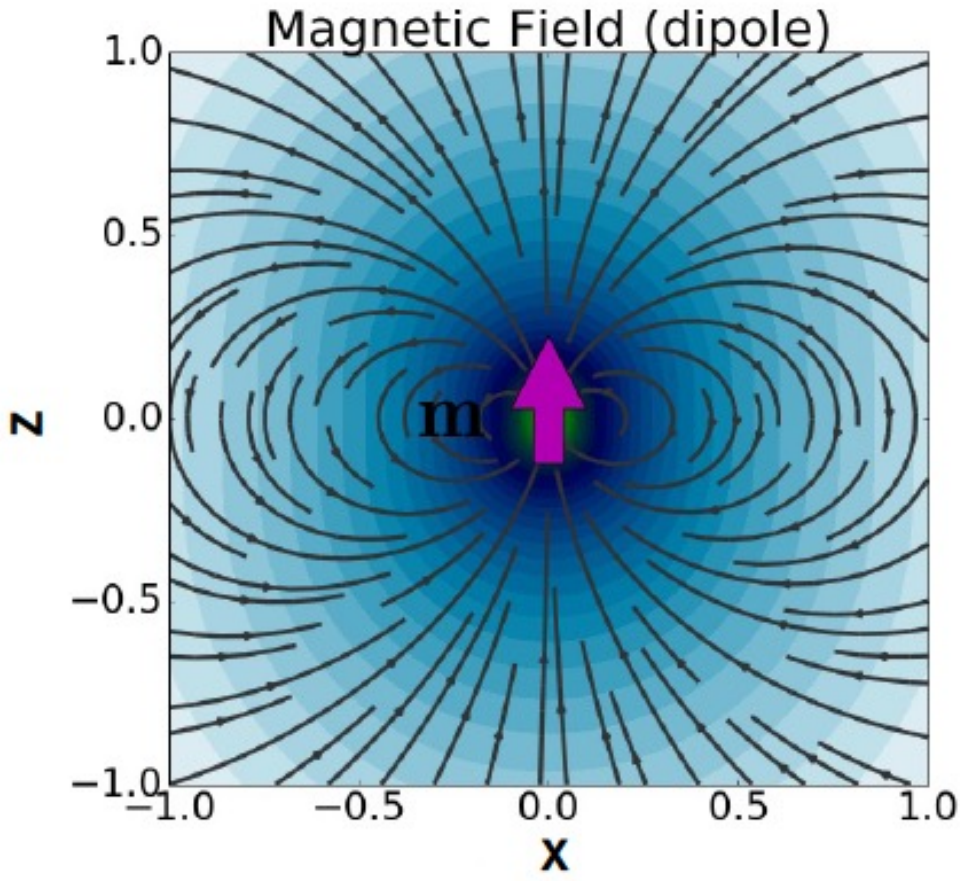}
\caption{A scale-independent lot of constant magnetic field lines of a point dipole along
$\hat{z}$, plotted in the $x$-$z$ plane.
The large arrow marks the position and direction of the dipole;
arrows indicate the direction of the field at that point. 
The darker blue shading indicates larger magnitudes of the $\vec B$ field,
which varies as $1/r^3$.}
\label{fig:dipole}
\end{figure}

{\it Magnetic disturbance.}
The conventional dipole vector potential and magnetic intensity 
for an isolated muon moment $\mu$ is provided in the SM.
Expressions will incorporate $\vec\mu=\mu (0,0,1)$
defining the $z$-direction)
\bea
\vec{A}^{\mu}(\vec r)&=& \nabla\times \frac{\vec\mu}{r}
                        =\frac {\vec{\mu}\times {\hat r}} {r^2}
                        = \frac{\mu}{r^3} (-y,x,0)\nonumber \\
\vec{B}_{tot}^{\mu}(\vec r)&=& \nabla\times\vec{A}^{\mu}(\vec r)
    = \frac { 3\hat{r} (\hat{r} \cdot \vec{\mu}) -\vec{\mu} } {r^3}
    +\frac{8\pi}{3}\mu\delta(\vec r)
\label{eqn:1}
\eea
with dipole $\vec B^{\mu}_{dip}$ and contact $\vec B^{\mu}_{con}$
terms. The divergenceless gauge is used for 
the vector potential.\cite{London1935} This form of magnetic field has 
amplitude $\mu/r^{3}$ times an angular term of order unity, 
$\vec A\sim \mu/r^{2}$ is more extended. 
Dipolar $\vec B(\vec r)$ field lines are
pictured in scale independent form in Fig.~\ref{fig:dipole}, 
Since symmetries will become central in the discussion, a symmetry
that is not obvious at first is that $\vec B(-\vec r)$=$\vec B(\vec r)$,
in addition to obvious axial symmetry.

\begin{figure}
 \includegraphics[width=0.6\linewidth]{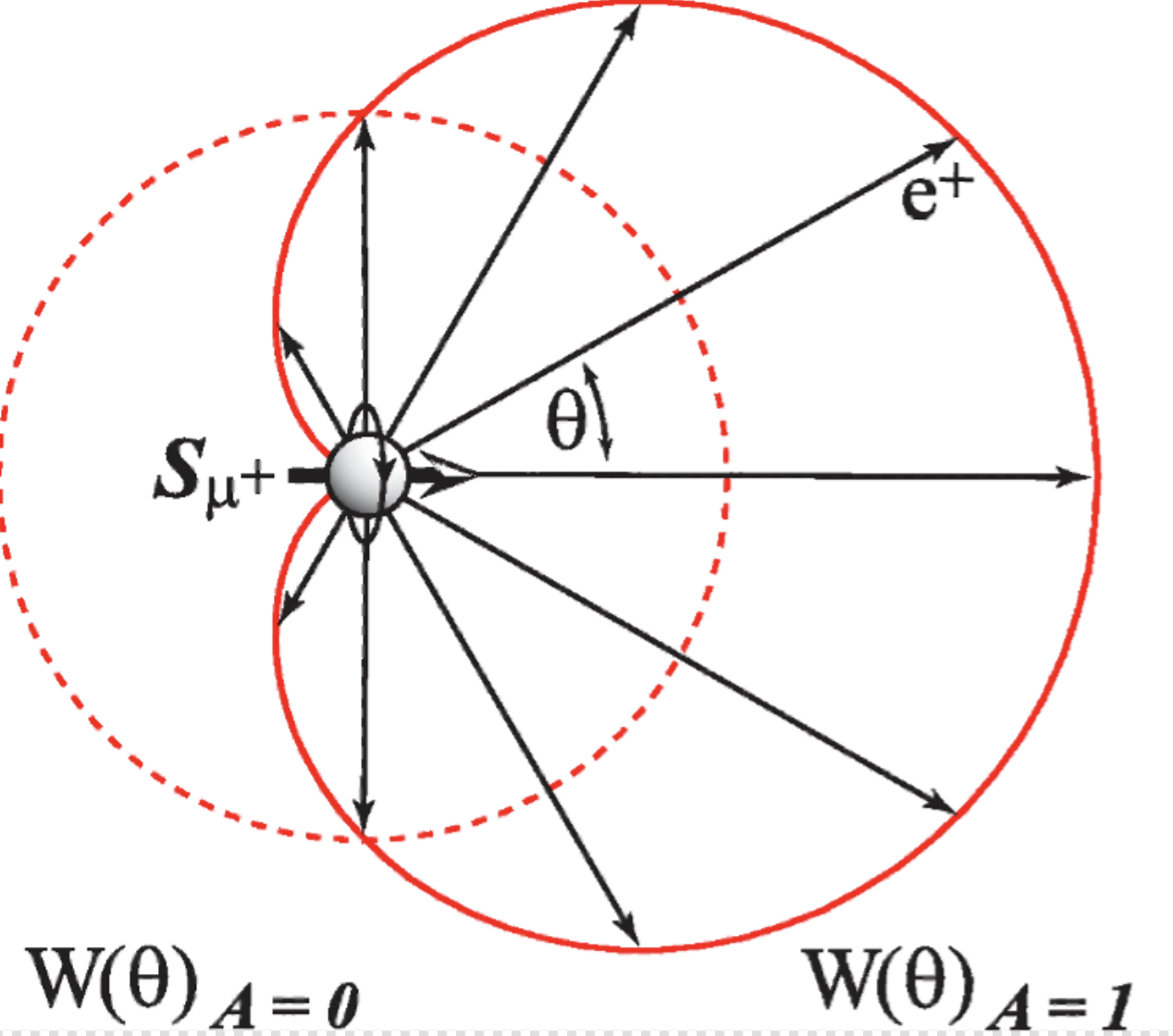}
\caption{For muon decay: the angular distribution of the
direction of emission of the decay positron that is detected. The
distribution is  $W(\theta)\propto 1+A\cos\theta$, with $\theta$
being the angle with respect to the muon moment (arrow) at the time
of decay. The factor $A$ depends on the energy of the emitted positron.
Two positron energies are pictured: dashed line, 25 MeV near minimum,
$A\approx 0$; full red line, 50 MeV near maximum, $A\approx 1$.
For energy scales: the rest mass of the muon
is $207\times 0.511$ MeV $\sim$ 100 MeV, which is what is 
available for the decay products, which include a neutrino and
an anti-neutrino.
}
\label{fig:PositronEmission}
\end{figure}

The muon is implanted with a polarized spin, {\it i.e.} the $z$-component
(chosen direction) has spin quantum $\sigma_z=+1/2$. The spin-half
moment includes $x$ and $y$ components of equal r.m.s magnitudes of
1/2, oriented in unknown direction. The decay of the positive muon,
with mean lifetime 2.2$\mu$s,
emits a positron, a neutrino and an antineutrino, with positron 
emitted along the (vector) spin direction; all particles have a
distribution of energies that satisfy energy-momentum conservation.
The distribution of the emitted positron depends on its energy, as
pictured in Fig.~\ref{fig:PositronEmission}, being spread over the 
positive hemisphere of solid angle centered on the muon.

Returning to Eq.~\ref{eqn:1}, it is conventional that in the midst
of a electronic system, the contact term in Eq.~\ref{eqn:1}
will give rise to a `contact field energy' 
\bea
 E_{\mu-el}^{cont}=-\frac{8\pi}{3} \vec{\mu} \cdot \vec m(0)
\eea
in terms of the electronic magnetization density $\vec m(\vec r)$.
This term enters hyperfine splittings which will not be a topic of
this paper.

{\it Charge disturbance.}
 The immediate local environment of an interstitial muon is that of 
a positive charge in a slowly varying electron gas, where it attracts one unit of 
electron density to its vicinity from the valence density.  A first 
{\it approximation} is that of the constant electron density (jellium) plus the 
attracted density into an effective electron $1s$ orbital, centered at the 
muon's site and spherically symmetric, given approximately by
 $n(r)$=$n_o$+$n^{\mu}_{1s}(r)$, where $n_o$ is the value of the 
uniform jellium density and $n_{1s}$ is an effective $\mu^+$$1s$ orbital density.
 Neutrality apart from decaying Friedel oscillations will be achieved 
within roughly the Thomas-Fermi screening length. 

At interstitial densities of densely packed intermetallic
 compounds characteristic of most SCs, hybridization of the muon
 $1s$ orbital with itinerant electronic
 states will ensure that the muon ``atomic density'' will not be spin
 polarized due to electron gas exchange effects. (Some DFT studies
indicate a muon bound state below the bottom of the conduction 
band,~\cite{Huddart2021}
in which case single occupation effects (Hubbard U) and magnetic polarization
effects would arise, with the $1s$ moment forming a singlet with the 
$\mu^+$ spin at zero temperature.) 
At this level the discussion is
 that of a proton in a HEG,\cite{Duff2007} except that
 the proton's vector potential is smaller and the $\mu^+$ quantum
position uncertainty is greater, as discussed later.

{\it Magnetic  polarization: spin.}
At a sufficient distance from the strong and diverging short range
field  near the muon, the induced
magnetization at position $\vec{r}$ is given from linear response as
 \begin{eqnarray}
 \vec{M}^{ind}(\vec{r})=
   \mu_B \big[n_{\uparrow}(\vec r)-n_{\downarrow}(\vec r)\big]
  \rightarrow \chi_{sp}[n(r)] \vec{B}^{\mu}(\vec{r}).
\label{eqn:Mind}
 \end{eqnarray}
The last expression includes the Pauli spin susceptibility $\chi_{sp}$,
with this linear response result commonly remaining realistic to 
multi-tesla-scale fields. [Recall that $\mu_B$$B$
at one tesla is 0.7 K in temperature units, which will be a factor of 207
smaller for the muon moment.]
 In this HEG approximation, which is formalized by density
functional theory,\cite{Janak1977,RMMartin}
 $\vec{M}^{ind}$ is parallel to $\vec{B}^{\mu}$ at each point,
hence it and the
entire system satisfies the same cylindrical and reflection symmetries given
for $\vec{B}^{\mu}$.
In a HEG the induced field
at the muon $\vec B^{ind}(0)$ will, by symmetry, align with the muon
moment. It will then not provide any torque on the muon moment, hence
be undetectable by depolarization studies. The induced field will rotate
with the muon spin as it precesses due to an applied or spontaneous field.  That
the muon lies at a low symmetry position is relevant and is discussed later.

{\it Orbital polarization.}
 The muon-induced electron currents will produce a local orbital 
polarization and related susceptibility
 $\chi_{orb}$. For the HEG and a uniform magnetic field the value 
of this (Landau diamagnetic) susceptibility is 
$\chi_{orb}=-(1/3)\chi_{sp}$. In the following discussion a
net paramagnetic susceptibility $\chi_p=\chi_{sp}+\chi_{orb}$ 
in the normal state will be
accounted for. Nonlinear effects (in the high field region) will be 
discussed separately. In normal density Fermi liquid metals, bulk 
$\chi_p$ is of the order of $10^{-3}-10^{-4}$, in $cgs$-$gaussian$ units. 
This small polarization may be relevant
because the inferred spontaneous field reported in these metals is 
unusually small.
For the muon's strongly non-uniform near-field,
 orbital effects in the normal state have reduced effect. 


{\it Induced magnetic field.}
 Each electron carries a magnetic moment of one $\mu_B$, thus
each volume element of induced magnetization 
 $\vec{M}^{ind}(\vec{r})\Delta V$ will produce
 the same form of magnetic field intensity from the incremental moment
 $\vec{M}^{ind}\Delta V$ as given by the usual dipole
expression, except that the initial origin $\vec{0}$ will be
assumed by $\vec{r}$ (the position of $\Delta V$) and the 
position of a given field point will be
$\vec{r'}$. Some details are presented in 
Appendix~\ref{app:inducedpolarization}. 

The net result of the polarization is the magnetic flux
density in the region (in $cgs$-$gaussian$ units)
\begin{eqnarray}
\vec B^{tot}(\vec r)&=&\vec B^{\mu}(\vec r) + \vec B^{ind}(\vec r)\nonumber \\
  &=&\vec B^{\mu}(\vec r)+4\pi\chi_p \vec B^{\mu}(\vec r) \nonumber \\
   &=& \big[1+4\pi\chi_p(\vec r)\big]\vec B^{\mu}(\vec r),
\label{eqn:3}
\end{eqnarray}
The result is a textbook-like result, reminding again that this local
linear response enhancement is good except within a small volume
surrounding the muon, to be addressed in the next section. The polarization response is
comparatively small, however, small magnetic fields are the topic of this paper.
 
Integrating the effect of the induced moment over all space, 
the magnetic field created by aligned vector potentials of the electrons
corresponding to $\vec M^{ind}(\vec r')$ will be 
 \begin{eqnarray}
 \vec{B}^{ind}(\vec{r'})&=&
    \int d^3r \frac {3\hat{R} [\vec{M}^{ind}(\vec{r})\cdot
      \hat{R}] - \vec{M}^{ind}(\vec{R})}  {|R|^3}. 
 \label{eqn:dipoleH}
 \end{eqnarray}
where $\vec R\equiv\vec{r'}-\vec r$.
Simplification occurs because we are only interested in the field at 
the muon site, {\it i.e.} at $\vec{r'} \rightarrow 0$, so $\hat{\vec{R}}
 \rightarrow -\hat{r}$, and that the source is a point (the muon).
Due to the symmetry of any dipolar field 
$\vec B(-\vec r)=\vec B(\vec r)$ the induced magnetization gives,
in the quasiclassical expression of Eq.~\ref{eqn:dipoleH}
arising from linear response, a divergent result at the muon site
due to neglected quantum effects, which is discussed in 
Appendix~\ref{subsec:fieldatsite}. The linear response assumed in Eq.
(\ref{eqn:3}), and therefore Eq.~(\ref{eqn:dipoleH}), breaks down
near the muon.
Quantum corrections are necessary to give regular results, see
Refs.~[\onlinecite{Slichter1996,Abragam2012,Autschbach2012}]
and Sec.~\ref{sec:QuantumEffects}. 
Necessity for quantum corrections to the supercurrent-generated 
magnetic field will arise in Sec.~\ref{sec:superstate}. 

\section{Muon near-field processes}
\label{sec:QuantumEffects}
\subsection{Quantum effects}
The near-field region of the integral for $B_z^{ind}(0)$
follows from Eq.~\ref{eqn:dipoleH}, with detail presented in
Appendix \ref{app:inducedpolarization}, 
involves more than non-linear corrections to the induced magnetization.

{\it Quantum positional uncertainty (QPU).}
QPU, commonly and imprecisely referred to as zero
point motion, contains information on how a confined quantum particle
samples a region around the classical position. In its ground state 
the muon will sample a region around a minimum of the Coulomb potential,
the minimum being the classical ground state position. Theoretical
description and implementation of QPU, and anharmonicity as well, has undergone
recent substantial progress.\cite{Monacelli2021,Errea2020,Errea2016}
This uncertainty is sometimes important for interstitial protons,\cite{DFT1} but the
effect will be larger for a muon with its factor of nine smaller mass,
 be less regular in shape\cite{Huddart2021} for a low symmetry environment,
and may even involve a quantum oscillation between two classically-preferred 
sites.\cite{Huddart2022}

Until recently, this QPU of the muon position had not become a
mainstay of $\mu$SR analysis. The first step, finding the classical 
muon position, has been made
more efficient through a user-friendly platform, since the advent of the
$\mu$SR analysis application MuFinder.\cite{Huddart2022} With only the underlying 
crystal structure as input, the algorithm chooses likely sites for the muon,
calculates from DFT the energy of the system including relaxation (structural and
electronic) of nearby atoms, and with calculated forces iterates to 
the minimum energy structure.
The results can be used in the analysis of $\mu$SR data, especially for
nuclear magnetic fields and for magnetic solids. 

Beyond this classical position of the muon, its QPU has now been
realized as important for the interpretation of $\mu$SR data. 
In the second step, the factor of 207 difference in muon and electron 
masses allows one to invoke an adiabatic approach as first approximation: 
for each point $\vec R$ within the muon ground state
normalized wavefunction $\Psi(\vec R)$, the electron density $n(\vec r;\vec R)$
and polarization 
(magnetization) $M^{ind}(\vec r;\vec R)$ can be evaluated, and from it the magnetic
flux density $B^{ind}(\vec r;\vec R)$. 
The resulting (``smeared'') induced  magnetic
field is given by the expectation value
\bea
\vec{\bar{B}}^{ind}(\vec r)=\int d^3R \Psi^*(\vec R) {\vec B}^{ind}(\vec r;\vec R)
                               \Psi(\vec R).
\eea
Here $\vec r$ is measured from the classical muon position $\vec R_o$.
This field is then evaluated at the muon minimum energy position.
Onuorah {\it et al.} have formalized this expression using their `double
Born-Oppenheimer approximation' approach,\cite{Onuorah2019} and applied it to various
elemental transition metals and binaries as an ensemble of `classical positions'
of the muon to provide the above integral. Given other near-field complications
that will be mentioned, the integral for $\vec {\bar{B}}(\vec r)$ has not yet
been attempted, but QPU will be necessary for a
realistic treatment of the near-field.

{\it Region of full polarization.} 
The diverging magnetic field at the muon site will completely spin polarize the 
nearby conduction electrons along the lines of its dipolar field, modulo 
quantum restrictions.  Specifically, inside some small radius the 
spin polarization will approach  100\% along the local direction of the field.
This situation suggests an analog of the study by Ortiz, Jones, and
Ceperley\cite{Ortiz1995} of the H$_2$ molecule in superstrong 
applied magnetic field.  
They applied quantum Monte Carlo (fixed phase and variational) methods
and determined, among other things, that the ground state is a
TRS-breaking electronic triplet, having angular momentum $L_z=-1$
relative to the H$_2$ axis. Note that these methods do not
presuppose atomic orbitals.  
Their study provides probable methods to study the muon problem
at radii where non-linear magnetism develops or, as we now discuss,
may become inhibited..  

When the polarization (here spin $s$=$\uparrow,\downarrow$ indicates the direction
with respect to the the local magnetic field), following the magnetic field shown in
Fig.~\ref{fig:dipole},
\bea
P(\vec r;B,T)=\frac {n_{\uparrow}(\vec r;B,T)-n_{\downarrow}(\vec r;B,T)}
                {n_{\uparrow}(\vec r;B,T)+n_{\downarrow}(\vec r;B,T)}
\label{eqn:paircorrelation}
\eea
approaches unity, and the analog of the Ortiz {\it et al.} condition of
triplet alignment may apply.\cite{Ortiz1995}  Here $n_s(\vec r);B,T)$
is the density of dynamically available electrons at temperature $T$ in
field $B$, consistent with fermion occupation.  Any additional longitudinal field 
can produces no extra  polarization. 
Thus the longitudinal magnetic susceptibility 
goes to zero as some power of the distance $r$ from the muon site.

{\it Electronic pair correlation.}
From variational~\cite{Duff2007,Pickett1993,Ortiz1994} and 
quantum~\cite{Spink2013} Monte Carlo calculations
on correlated wave functions (Gutzwiller-Slater determinants)
at full polarization, the probability of parallel spin electrons (all
electrons in this limit) being at the same point vanishes (Pauli repulsion).
The pair correlation function in three dimensions increases from zero
quadratically, further reducing polarization in the near-field region. With
the other two effects, the near-field will present no conceptual difficulty
but leaves a challenge in predicting the field at the muon site.

\subsection{Importance of anisotropy}
\label{subsec:anisotropy}
Since the interstitial muon site usually involves little or no
symmetry (due to relaxation of neighboring atoms), anisotropy of the 
muon environment must be confronted. The actual symmetry experienced
by the muon will be low.\cite{Bernardini2013,Bonfa2015,Huddart2022},
Considering the crystal symmetry,
for a space group with N operations and a site without symmetry, there
will be N symmetry related sites in the unit cell but with different
orientations of the local environment with respect to the fixed muon 
polarization, hence different projections of a
muon-induced field along the muon polarization axis. This is a
material dependent, symmetry dependent task to address.
It follows that the induced magnetization and its magnetic field 
will no longer have the symmetry of the point dipole field, thus 
the field at the muon site at a given time will not align with
the muon spin, thus will produce a torque on the muon moment, with
different torques for the symmetry-equivalent muon sites.
The effective field at a given
muon site may become amenable to DFT+ studies,\cite{Kuster2021}
(+ indicates manybody or quantum corrections).
What seems clear is that it
is necessary for quantitative studies to (i) determine the classical site of the muon,
(ii) calculate the quantum uncertainty of the muon position, which may 
adjust the muon's most probable position, (iii) account for the motion
of the muon during its lifetime, then (iv)
determine the resulting effective field near and at the muon site taking
into account these various effects.


\section{Reports of TRSB}
\label{sec:TRSB}

\subsection{Reports as of this writing}
\label{subsec:TRSBlist}
Fairly early on superconducting states, especially as probed by
$\mu$SR, were proposed as a platform for observing TRSB.\cite{Kiefl1990}
Before considering the muon+superconductor system, the following list of 
reports is provided, as this is the point of this paper.
The following near comprehensive list (as of this writing) includes
a variety of materials types, space groups, and degree of anisotropy, 
each followed by the reported spontaneous field, 
or by NA if the field was not available. All have Fermi liquid normal
states, with low T$_c$ values consistent with weak coupling.\\
$\bullet$ skutterudite Pr(Os,Ru)$_4$Sb$_{12}$ 
      (0.6 G)\cite{Ghosh2021,Aoki2003}\\
$\bullet$ non-centrosymmetric Re$_6$Zr (0.6 G, 
      1.2 G)\cite{Ghosh2021,Singh2014}\\
$\bullet$ non-centrosymmetric LaNiC$_2$ (0.1G)\cite{Hillier2009},
        (NA)\cite{Quintanilla2010,Sumiyama2015,Sundar2021}\\
$\bullet$ Dirac silicides (Nb,Ta)OsSi (0.83 G, 0.17 G)\cite{Ghosh2022}\\
$\bullet$ quasi-skutterudite Lu$_3$Os$_4$Ge$_{13}$ (1.1 G),\cite{Kataria2023a}\\
$\bullet$ putative frustrated superconductor Re$_2$Hf (1.2 G),\cite{Mandal2022}\\
$\bullet$ non-centrosymmetric intermetallic La$_7$Pd$_3$ (NA)\cite{Mayoh2021}\\
$\bullet$ rocksalt structure monosilicide ScS (NA)\cite{Arushi2022}\\
$\bullet$ centrosymmetric LaNiGa$_2$ (NA)\cite{Hillier2012},
         (0.2 G).\cite{Ghosh2021}\\
$\bullet$ cubic Re$_6$(Zr,Hf,Ti) (0.2 G)\cite{Ghosh2021}\\
$\bullet$ unusual $I4/mmm$ tetragonal SC Sr$_2$RuO$_4$ (0.5 G)\cite{Ghosh2021}\\
$\bullet$ tetragonal Fe-based Ba$_{1-x}$K$_x$Fe$_2$As$_2$ (0.1 G)\cite{Ghosh2021}\\
$\bullet$ non-centrosymmetric hexagonal SrPtAs (0.07 G)\cite{Ghosh2021}\\
$\bullet$ non-centrosymmetric tetragonal CaPtAs (0.8 G)\cite{Ghosh2021}\\
$\bullet$ cubic (bcc) Re$_{0.82}$Nb$_{0.18}$ (0.4 G)\cite{Ghosh2021}\\
$\bullet$ hexagonal $P6_3/mmc$ Re (0.2 G)\cite{Kataria2023,Ghosh2021}\\
$\bullet$ non-centrosymmetric  La$_7$(Ir,Rh)$_3$ (0.1 G)\cite{Ghosh2021}\\
$\bullet$ tetragonal $I{\bar 4}2m$ Zr$_3$Ir (0.08 G)\cite{Ghosh2021}\\
$\bullet$ tetragonal $I4_1/acd$ (Lu,Y,Sc)Rh$_6$Sn$_{18}$ (0.6 G)\cite{Ghosh2021}\\
$\bullet$ bct CaPd$_2$Ge$_2$, T$_c$=1.7K (0.34G)\cite{Anand2023}\\
$\bullet$ skutterude-like Y$_3$Ru$_4$Ge$_{13}$, T$_c$=2.8K (1.8G)\cite{Kataria2023}

More than half of these examples are ternary or quaternary compounds, only one is 
an element (hexagonal Re). Due to this and the observation that several 
Re$_{1-x}$$T$$_x$ alloys, $T$=transition metal, have also shown TRSB signals,
the authors of this report\cite{Shang2018} suggested that the local electronic
structure of the Re atom may be responsible for the signal. 

Although this paper concerns primarily $\mu$SR experiments, separate
indications of TRSB signals have appeared.
Of the examples provided above, Sr$_2$RuO$_4$\cite{Xia2006} and 
PrOs$_4$Sb$_{12}$\cite{Levenson2018} have been proposed
as TRSB by observation of Kerr polar effect (KPE) rotation angles of
0.06 $\mu$rad and 0.25 $\mu$rad respectively,
and for LaNiC$_2$ by an ambitious direct magnetization measurement, 
reporting a field of 0.01 G.  
(See Sec.~\ref{sec:summary} for more discussion of Sr$_2$RuO$_4$.) 
Ghosh {\it et al.}\cite{Ghosh2021} provide references to the original
papers for the field values they provide. In a few cases, such as 
Re$_6$Zr, analysis shows enhanced depolarization below T$_c$ when
analyzed with Lorentzian relaxation but not with Gaussian relaxation, or
vice versa. Table 2 of the Ghosh {\it et al.} paper\cite{Ghosh2021}
 provides a list of correlated electron materials that have been probed 
with ZF-$\mu$SR or PKE, some showing a TRSB signal and some not.
Section~\ref{sec:sample} provides examples where there have been
conflicting reports of TRSB, or not.
For LaNiGa$_2$, a 2026 NMR report on 
spin-lattice relaxation 1/$T_1 T$ found a pronounced
Hebel-Slichter peak \cite{Curro2026} below T$_c$. Such behavior is
considered to be one of the most definitive indications
of singlet pairing. 

Discussion of some aspects of depolarization analysis is
provided in Sec.~\ref{subsec:analysis}.
The signal of depolarization as T$\rightarrow$0 is a few to several 
percent of the normal state value,
sometimes with an anticipated T-dependence (qualitatively like the gap) 
but sometimes appearing somewhat below
T$_c$ and sometimes with unclear T-dependence (due to uncertainty in data
or sparse temperature measurements).  
As an example, for Re$_6$Zr\cite{Ghosh2021} with T$_c$=6.8 K 
(much the highest T$_c$ in the above list of SCs), 
the maximum signal at T$\rightarrow$0 is 3\% above the
normal state value, with the associated spontaneous field value having been
provided as either 0.6 G or 1.2 G, not a serious discrepancy considering
that these values are near the limit of detection.  For LaNiGa$_2$,
the sample case we have chosen for this article, with T$_c$=2 K, the increase
is 6\% above the normal state value as $T\rightarrow 0$ and is representative
of that of a secondary OP.\cite{Hillier2012}
See Sec.~\ref{sec:sample} for some discussion of sample dependence.

\subsection{Comments: symmetry breaking,  order parameter}

The SC state encountered by the muon is addressed in the 
following sections, with discussion of the order parameter 
in Sec.~\ref{sec:OP}. As an indication of the issues
given the experimental indication of TRSB, why does the proposed exotic
state of weakly coupled, Fermi liquid metals, viz. LaNiGa$_2$, 
require a combination of \\
$\bullet$ breaking of U(1) symmetry [a given]\\
$\bullet$ avoidance of isotropic singlet spin pairing [uncommon]\\ 
$\bullet$ change of magnetic symmetry (TRSB) [uncommon]\\
$\bullet$ broken orbital/band symmetries at T$_c$ [uncommon]\\ 
and in some cases (viz. LaNiGa$_2$) should do so in \\
$\bullet$ a nonunitary manner [uncommon]\\ 
One `given' and four 'uncommons' multiply to
a `highly uncommon' occurrence -- a rare sighting. The topological band 
character of LaNiGa$_2$, due to two Dirac points on the Fermi surface, 
does give a distinction from
the other cases listed in Sec.~\ref{subsec:TRSBlist}, but several of the 
above questions extend to other 
members of this class of fragile magnetic superconductors.

A primary question, addressed by several theory groups pursuing exotic
pairing (many cited in the Introduction), is: 
what type of order parameter is consistent with the 
measured constraints, and precisely what are the constraints? 
It has been understood since not long after BCS theory appeared
that a magnetic impurity diminishes a singlet order parameter locally 
(by providing pair-breaking) and
degrades SC properties. The muon is a magnetic impurity.
A triplet OP will respond differently from a singlet state, in
ways that depend on the form of the triplet OP.~\cite{Rosenstein2015}
The following sections deal with the electronic behavior underlying
these questions, then addressing them to some extent.

\section{Superconducting state}
\label{sec:superstate}
\subsection{Supercurrent and its effect}
\label{subsec:supercurrent}

\subsubsection{Supercurrent response to the vector potential}
In identification of a bulk superconducting state, more convincing
than zero measured resistance is the magnetic susceptibility, reflecting
the transition from paramagnetic to dielectric character. The magnitude
of the diamagnetic susceptibility below T$_c$ provides, among more
detailed inferences, the superconducting fraction of the sample. 
A fundamental property is the response of the SC material to the
imposition of a magnetic field, {\it i.e.} spin-antiparallel Cooper
pairing versus spin-parallel alignment of spins with the magnetic field.
The triplet case should provide low energy pair spin flips
without depairing, hence a 
paramagnetic susceptibility, versus the large diamagnetic susceptibility
of singlet SCs.

The theory of the origin of the supercurrent has gone through 
progression as the theory of SC has been fleshed out. A very brief accounting
of the various expressions\cite{London1935,Ginzburg1950,BCS1957}     
relating the vector potential to the magnetic field is 
given in Appendix~\ref{app:3theories}, with more discussion in
Ref.~[\onlinecite{superfluid}].
Each gives a result in the London form,\cite{London1935} a staple of
superconductivity theory,
\bea
\vec J^s(\vec r,T)=-\frac{1}{\Lambda(T)}\vec A^{\mu}(\vec r),
\label{eqn:Londoneqn}
\eea
where $\Lambda$ incorporates the linear (and local) response of the superconductor. The
three expressions differ, especially the latter (BCS), but do not affect the 
discussion in this paper. 

The point of interest here is more general: a magnetic field in a
superconductor interacts with the charge directly through the vector potential
$\vec A$, not through the magnetic field $\nabla\times\vec A$. 
While this is a fundamental
precept of SC theory and phenomena, the interaction of the moment with
the SC order parameter has never, to the knowledge of this author,
been theorized on the small distance scale of the muon vector potential.
Given Eq.~\ref{eqn:Londoneqn}, $\vec A^{\mu}$ creates a supercurrent 
proportional\cite{Ashcroft2020} to $\vec A^{\mu}$,
as required by the Meissner effect. 

However, in a region where the muon field 
exceeds the experimental upper critical field, superconductivy will 
be quenched and there are other considerations. Gor'kov's Greens function 
theory,\cite{Gorkov1959} which
verified the Ginzburg-Landau theory below and near T$_c$, may be
generalizable to address short length scale behavior.
The scale on which this happens -- varying by orders of magnitude
over a few Angstroms -- is 
less than the penetration depth or coherence length.
This length scale is not addressed by the theoretical progression
mentioned above. The Usadel Green's function approach,\cite{Usadel} in which 
the SC gap 2$\Delta(\vec r)$ becomes position dependent due to disorder, a
magnetic field, or proximity to a boundary,
may be adaptable to the case of a position dependent, short range
vector potential such as that provided by the muon.

\subsubsection{Field at the muon site}
\label{subsec:fieldatsite}
The supercurrent density $J^s$ with circular form from $\vec A^{\mu}$ 
produces a magnetic field $\vec B^{sc}(\vec r)$.  Each $z$ level
(a circular sheet of current loops)
 produces a contribution to $\vec B^{sc}(\vec r)$
without any cancellation.
The muon experiences the combined field (not its own)
\bea
  \vec B^{tot}(0)=\vec B^{ind}(0)+\vec B^{sc}(0),
\eea
the last field from the supercurrent
arising only below T$_c$. The  induced field $\vec B^{ind}$ will 
reduce and vanish as the gap opens.

We focus on the vector potential $\vec A^{\mu}$, which leads to
a field arising from the induced spin magnetization $\vec B^{ind}$ and 
to the field arising from the supercurrent $\vec B^{sc}\propto \vec A^{\mu}$. 
Theory of a superfluid region at this sub-nanoscale is unavailable, so
applying the conventional theory 
is what is available. 
From magnetostatics, the field due to a (super)current, and its value
at the muon site, is (keeping only $\vec B^{sc}$ as the strongest
field)
\bea
\vec B^{sc}(\vec r)&=&\frac{1}{c} \int \vec J^s(\vec r')
          \times\frac{\vec r -\vec r'}{|\vec r -\vec r'|^3}\nonumber \\
   \vec B^{sc}(0)&\propto& \int \vec A^{\mu}(\vec r')
     \times\frac{\vec r'}{|\vec r'|^3}d^3r' \nonumber \\
    &\propto& \mu {\hat z}\int \frac{\rho^2}{(\rho^2+z^2)^{3/2}}d^3r
\eea4
where cylindrical coordinates $\rho, \phi, z$ for $\vec r'$ 
have been introduced, and using 
$\vec A^{\mu}=(\vec\mu \times \vec r)_/r^3$, $\vec\mu=(0,0,\mu)$.

The integral for the $x$ and $y$ components vanishes, consistent
with the symmetry for a homogeneous environment allowing 
only a $z$ component. (For an anisotropic environment these
components will not vanish.) The integral becomes
\bea
B^{sc}_z(0)&=&-\frac{4\pi c\mu}{\lambda_L^2} 
      \int_o^{\infty} \rho d\rho~\int_{-\infty}^{\infty}dz
          \frac{\rho^2}{(\rho^2 + z^2)^{3/2}}.
\eea
The integrand diverges in the limit $\rho\rightarrow 0$ or 
$z\rightarrow 0$. In this SC case there is however a
natural cutoff: when the field $B^{\mu}$ reaches $H_{c2}$, 
$|\vec B^{\mu}(\vec r_c)|=H_{c2}$, the region inside $\vec r_c$ will
be normal. This cuts off the integral, leaving the near-field behavior
within this region with quantum correction as discussed earlier,
if needed.

For the homogeneous electron gas with cylindrical symmetry of
the system, the field at the muon site from the supercurrent
will align with the muon moment
thereby providing no torque and no depolarization. Again, the low
symmetry of the muon site will cause the field to deviate from the
$z$ direction, thereby providing torque on the muon moment and
possibly a mechanism of depolarization.

\subsection{Kondo physics; YSR states}
\label{subsec:YSR}

\subsubsection{Magnetic moment coupled to pairing}
The well understood Kondo picture of a magnetic impurity in an electron gas 
addresses the coupling of the spin degree
of freedom of a magnetic impurity to itinerant electrons through
an on-site interaction $J_K \sum_{j}\vec S\cdot \vec s_{j}$ in
terms of a spin-electron Kondo exchange parameter $J_K$ and the 
impurity and electron ($j$) spin operators $\vec S, \vec s_j$, respectively.
At low temperature low-energy processes at the Fermi surface lead to
a ``local singlet,'' {\i.e.} a local screening of the impurity moment.
The muon brings also a unit charge that attracts a compensating electronic
charge within the screening length of the electronic system, typically
just a few Angstroms. Thus in the normal state a single interstitial $\mu^+$
has moderate effects, as discussed earlier.
 
As Cooper pairing initiates the picture changes. First, low energy
excitations below $2\Delta(T)$ are frozen out, tending to nullify Kondo singlet 
formation. Secondly, an itinerant electron in the area is frustrated 
between anti-aligning with another electron (for singlet pairing), or with 
anti-aligning with the impurity spin, causing fluctuation. This 
picture for an isolated moment is relevant here, whereas for a Kondo
lattice a separate state of Kondo superconductivity may arise, which
is not relevant here.
Thirdly, a magnetic impurity imposes a vector potential with an orbital 
coupling in the SC state. A dissipationless supercurrent arises directly
proportional to the vector potential (the Londons' ``theorem''),
which provides its magnetic field, as discussed above.

Another concern arises from the $3\mu/r^3$ scaling of the muon magnetic field,
whose effect in the SC state apparently has not been established; the common relevant 
length scales are the coherence length $\xi$ and field penetration depth
$\lambda$ with values from a few nm to a few $\mu$m, rather than
a few interatomic distances.
When this field exceeds the upper critical field, the order parameter vanishes,
although OP behavior is yet to be established on such a small
scale. For the critical field of LaNiGa$_2$, this distance is 1.3\AA.
The muon then will always exist in a small normal state region. 
This change from bulk order parameter to a normal region occurs on a smaller
length scale than $\xi$ (the small length scale in Type II SCs,
determined from singlet pairing considerations), 
which brings up the (possibly) unexplored question of how this
crossover occurs on the \AA~scale. 

For comparison, scanning tunneling microscopy spectra
and S-I-S tunneling characteristics demonstrate that the OP extends 
to the surface, or to an interface and somewhat beyond, where there 
is a rapid drop from full amplitude but with sufficient overlap that
Josephson coupling survives. How the increasingly 
large muon near-field might affect a muon's detection of relaxation-inducing 
of a  self-induced field, and indeed the reaction of the OP
to such a rapidly varying field, are areas yet to be clarified,
with relevant questions raised 
 in Sec.~\ref{subsec:fieldatsite}. 
Miyake and Tsuruta explored the effect of the muon's magnetic field in a
$p$-wave superconductor (without accounting for the degenerate Fermi
liquid aspect of superconductors),
resulting in $\sim$1 G fields as an
estimation of the consequences.\cite{Miyake2017} 

\subsubsection{Kondo singlet versus Cooper singlet}
An early approach to this coupled local moment-pairing order
parameter issue is embodied in the
1970 results of Zittartz and M\"uller-Hartman,\cite{Zittartz1970} (ZMH)
who extended earlier works at the model Hamiltonian level that had
been influential in understanding the normal state Kondo effect. 
One viewpoint was that the opening of the SC gap interrupts the 
essential low energy Kondo physics, and the
quantum aspects of the spin could be treated less explicitly.

Appending the Kondo Hamiltonian with a BCS `pairing potential' term,
ZMH established within this model that as the gap opens, bound states
involving the local moment appear within the gap but near
the gap edges -- the YSR states, as foreseen by the YSR 
authors,\cite{Yu1965,Shiba1965,Rusinov1969,Rusinov1969b}) 
discussed below.  The character of these states should include
strong local moment character, and more recent studies have
clarified both their energetic and orbital character. The
next two subsections discuss these states.

According to this level of theory, a single muon impurity 
couples to the electron cloud that is beginning to pair into
Cooper singlets but also is encouraged to form competing Kondo singlets. 
The outcome becomes,
besides other spectral changes,  a pair of 
magnetic moment-derived localized bound states within the
gap, at energies near the gap edges at $\pm\Delta(T)$. This
connection provides a mechanism of coupling of the SC order
parameter to an impurity spin.
As derived in Sec.~\ref{subsec:fieldatsite} (not considering 
Kondo coupling), the $\mu^+$ moment
creates a spin polarized region around the muon 
that is changing direction with polar angle $\theta$, with
near-field greater than the critical field.
This electronic magnetization near the muon, having strong polarization
(varying with direction)
acts to obviate singlet SC pairing in that region (not
necessarily encouraging triplet pairing),
thereby reducing the gap magnitude, to the extent that any OP
can be treated on the Angstr\"om scale. The magnetic ion's (muon's)
field at some point dominates, providing a moment-quasiparticle coupling of
as yet unstudied character. Study of the SC-Kondo model has since that time been
addressed by more recent many-body techniques, see for example
Sykora and Meng's calculation of the interplay between the Kondo singlet
state and the induced YSR states.\cite{Sykora2022}
Some results of the study by Choi and Muzikar\cite{Choi1989} of a Kondo
impurity in an exotic superconductor are discussed in 
Appendix~\ref{app:Kondo-exotic}.

\subsubsection{Yu-Shiba-Rusinov states; $\mu$YSR states}
As mentioned, a magnetic impurity in a gapped singlet SC interferes with the 
OP, leading to bound
YSR (Yu-Shiba-Rusinov\cite{Yu1965,Shiba1965,Rusinov1969,Rusinov1969b})
states within the SC gap $2\Delta$ but, according to early model studies,
 near the gap edges at
$\pm \Delta$. Such states are analogous to shallow donor and acceptor states
in semiconductors, and might someday play a similar role in SC
electronics.  After the Kondo effect was elucidated, 
one viewpoint was that the SC gap
inhibited low energy processes (which dominate the Kondo effect) as being frozen
out, and the moment could be considered as classical.  As theoretical 
interest and experimental capabilities have progressed, quantum
behavior of the impurity spin has become a topic of more detailed study, while the
internal structure of the state raises interest. This area
of study should be regarded as ongoing, with relevance to this article
to be determined. The status of impurity-induced states in superconductors
as of 2006 was reviewed by Balatsky,
Vekhter, and Zhu.\cite{Balatsky2006}

Previous studies treated the impurity moment-electron
coupling, implicitly (point contact coupling) in early work but increasingly explicitly 
in more recent studies, as interatomic electronic 
exchange akin to (but more broadly than)
the rules of Goodenough,\cite{Goodenough} Kanamori,
\cite{Kanamori} and Anderson\cite{Anderson} between the impurity orbital
mixing with orbitals of itinerant states. 
This direction of study neglected the 
smaller effect of the orbital behavior near the dipolar field of the 
moment. In studying the muonic impurity at the atomic level, both effects
may become relevant. The environment in $\mu$SR studies provides a 
direct coupling between the muon vector potential (orbital effects) and
the SC condensate as well as creation of $\mu$YSR states. Coupling of 
the muon is of a distinct character due to the fact that it resides at
an interstitial site and the coupling, at the basic level, is through 
the vector potential. The effect is a
polarization of the muon $1s$ electron orbital,
rather than through (say, $3d$) atomic orbitals. Hydrogen interstitials
should introduce similar formal considerations, but with a
smaller effect and less specific probes.

As noted, the implanted muon is a close cousin of interstitial hydrogen. 
Given a more-or-less singly occupied $1s$ 
orbital  of the muon, typically lying within the valence bands of the
host, there is no anticipation of an electronic (spin) moment 
and a $1s$ orbital has no orbital moment. Regarding this point:
it seems that DFT studies of proton and muon interstitials\cite{
Bernardini2013,Bonfa2015,Huddart2021,Gomilsek2023,Blundell2023} have rarely
searched for spin polarized activity of the muon. 

Interest in an
``ultra-deep donor'' state in ultra-pure semiconductors received a boost from a
simple type of correlated electron calculation (a hands-on DFT+$U$
treatment formalized later) applied to the H $1s$
orbital in Ge, which identified such a spin-polarized state a few eV
(note: \underline{not} a few meV) below the gap in Ge,\cite{Pickett1977}
well into the Ge valence band. Screening in metals alters the
physics of such states. Back to the main topic: the muon vector potential field
will interact with the SC OP and it can be expected to give rise to defect
states within the gap, which will here be denoted $\mu$YSR states.
A study of persistent YSR resonances in an exotic SC state was presented
by Senkpiel {\it et al.}\cite{Senkpiel2019}

The study of YSR states has progressed to DFT studies of $3d$ ions with
selected symmetries.
Fe, with its conventional large moment, has been the preferred magnetic atom, within 
or on the surface, of conventional SCs
such as Pb\cite{Ji2008} and Nb,\cite{Saunderson2020} revealing a great amount of
detail that can occur in such cases. Some cases are addressed
in Appendix~\ref{app:realYSRstates}. In-gap states due to a muon, with only
the moment's vector potential and with
the unpolarized $1s$ orbital not involved, have yet to be studied, although
a few groups may soon have the capability (see App.~\ref{app:realYSRstates}).

\section{Implications for pairing symmetry}
\label{sec:OP}
Construction of plausible exotic order parameters relies on
the guiding principles of antisymmetry of the Cooper pair and
 the BCS form of the nonlinear gap equation,
which becomes linear at T$_c$.
For metals that display conventional Fermi liquid behavior without any
unusual magnetic tendencies, superconductivity is initially assumed and
then (frequently) verified to be due to phonon-induced pairing of electrons -- 
singlet $s$-wave pairing. When 
properties, in the normal or SC state, are unexpected, an exotic ({\i.e.} non-BCS)
OP is pursued. 

The two subclasses of
OP are spin-singlet and -triplet pairing, while each is
subject to orbital (real space, $k$ space) effects including symmetry reduction. 
TRSB signals (viz. magnetic) provide the dominant triplet viewpoint.
Beyond that, broken symmetry in momentum space (`orbital symmetry' of the
gap $\Delta_k$ beyond $s$-wave) provides the
second consideration. Accepting TRSB, spin (or orbital) magnetization becomes the
issue, arising from an accompanying symmetry breaking. This section provides 
an overview of the progression  of theory since the report of TRSB in
LaNiC$_2$.  
The reviews cited in the Introduction should be consulted for
a more nuanced discussion.

\subsection{The spin singlet issue}

Occam's razor\cite{occam} encourages one to shave away unduly involved models while
continuing to connect with observations. This maxim is of course not such
a straightforward and easy guideline to apply when faced with unusual data.
Pairing due to phonon glue should be considered first, as it provides the
presumption underlying experimental identification of materials properties
that lie within familiar ranges for weak coupling Fermi liquid superconductors. 
Cooper's demonstration
that the Fermi surface is unstable to formation of bound {\it singlet} pairs of
zero total momentum is the underpinning of the theory of
superconductivity, subject only to the necessity of a net-attractive
effective interaction. The BCS choice was exchange of a boson but
more specifically a phonon, because
phonons are always present and the isotope effect was known. 
Phonons are known to favor zero momentum singlet pairing,
subject to interruptions from exotic pairing
mechanisms that are negligible in conventional weakly coupled Fermi liquids.
Spin triplet scenarios have not relied significantly on the question of energy 
loss versus gain of the spin pairing choices, focusing instead on triplet
pairing with a magnetic component of the OP in conjunction with available 
experimental data.

\subsubsection{Ubiquitous el-ph coupling; Eliashberg}
\label{subsubsec:EPC}
Electron-phonon coupling (EPC) is distinctive given that it (i) is always 
in play, and (ii) is always attractive for $s$-wave pairing (the kernel in the 
gap equation does not change sign), with coupling strength
$\lambda$ moderated by a retarded Coulomb repulsion $\mu^*$, typically
falling in the 0.10-0.16 range.\cite{Bogoliubov1958,Morel1962} 
For low T$_c$'s, say below 5K, it has been difficult but is recently 
becoming possible due to extended methods, to verify 
theoretically\cite{Sanna2020,Pickett2023} 
that SC derives from EPC. 

The challenge is that calculating low T$_c$ 
requires {\it precise} knowledge of both $\lambda$ and $\mu^*$, 
especially the latter of which is
normally unavailable.  An added complication is that the fundamental Coulomb 
repulsion is $\mu$, which is very involved to calculate precisely; almost
all calculations of $\mu$ involve (possibly reasonable) approximations, but as 
EPC becomes weak, the precision of $\mu$ becomes important.  
$\mu^*$, used in the majority of solutions of the Eliashberg equations to date, and in
existing T$_c$ equations, is a numerical simplification that depends on 
energy scales and a frequency cutoff $\omega_{cut}$ used in a more
efficient solution of the Eliashberg equations,\cite{AllenDynes1975} so
$mu^*(\omega_{cut}$ is a computational convenience, not an observable. 

Taken together (they are each always present),
the combination $\lambda_k-\mu^*$ contains some richness, based on the $k$-dependence
of $\lambda_k$ and the near-independence of $\mu^*$ on wavevector (it involves
higher energy virtual processes that serve to average out the dependence on $k$ 
across the Fermi surface). If $\lambda_k$ is strongly anisotropic 
or $\mu^*$ is unusually large, the net
interaction $\lambda_k-\mu^*$ can change sign on the Fermi surface,
and non-$s$-wave orbital behaviors become possible.\cite{Foulkes1977}  
For a weak but anisotropic $\lambda_k$, a change in sign of
$\lambda_k-\mu^*$ in the kernel introduces the possibility of
a exotic gap symmetry, denoted $p$-wave, $d$-wave, or even more
involved combinations of symmetries, such as $s+id$. While a few $p$-wave and
$d$-wave SCs have been suggested for quantum materials, some with
several types of support, none has been established in
the weakly-correlated Fermi liquid metals that are the topic of this article.
Simple Fermi liquid SCs, with conventional $\mu^*$ and low T$_c$ are commonly
and successfully interpreted
to arise from EPC, due to the isotope effect on T$_c$ when available, 
partly due to other experimental information on $\lambda$,
and partly due to the ubiquity of the phonon mechanism.

A pairing strength $\lambda$=
0.45-0.6 in an $s$-$p$ electron compound with a standard frequency 
spectrum can account for a T$_c$ up to
5K or so (but is still sensitive to $\mu^*$), a range including all TRSB SCs
in the class of fragile magnetic superconductors. Given the
seemingly necessary exotic order parameter, an additional channel for
symmetry-breaking is anticipated.
While weak for EPC, this value of coupling $\lambda_k$ may be strong compared to
additional candidates, hence alternatives are more plausible
if they are complementary to,
rather than competitive with, EPC. More visually expressed, they
might be simply additional to EPC. 

Subedi and Singh~\cite{Subedi2009} calculated
the phonon spectrum, the Eliashberg spectral function $\alpha^2F(\omega)$,
and T$_c$ for LaNiC$_2$, a sister compound to LaNiGa$_2$
with closely related composition, filled Ni $3d$ bands, an orthorhombic
space group, and similar value of T$_c$=2.7K. It has a different
point group, but one that also has only 1D irreducible representations,
{\it i.e.} no degenerate crystal symmetry to be lifted.  The
$\mu$SR identification as TRSB is based on extraction from analysis of a
spontaneous field as small as 0.1 G (see Sec.~\ref{subsec:TRSBlist}), 
a value frequently stated 
as the lower limit of detectability. The calculated coupling strength 
is $\lambda$=0.52, and using a standard
value of the retarded Coulomb repulsion $\mu^*$=0.12, T$_c$=3K was obtained.
This value is consistent with experiment, and provides the strength of
electron-phonon
coupling that must be confronted by competing pairing mechanisms.

Related calculations on LaNiGa$_2$ were reported by T\"ut\"unc\"u and
Srivastava,\cite{Tutuncu2014} who assumed the $Cmmm$ space 
group\cite{Grin1982} understood
at the time to determine the structure, rather than the more recently 
discovered $Cmcm$ structure from single crystal XRD.\cite{Staab2022,Badger2022} 
The electronic structure is similar to the $Cmcm$
result, for example the Ni $3d$ states are filled and lie in the same
energy range, and given the similarities of the crystal structures 
the phonon spectra should be similar. The Fermi surfaces for both 
are large and multisheeted, but
different. Based on a limited $Q$-mesh for the phonons, they obtained
$\lambda$$\approx$0.7, and choosing $\mu^*$=0.17 for their estimate, 
their calculated $T_c$ was
close to the 2 K experimental value. Calculations using the
more recently determined $Cmcm$ space group have not been reported.

Calculated values of T$_c$, when small, are sensitive to (i) the 
choice of the parameter $\mu^*$, often treated as a parameter
but reflecting the retarded nature of the Coulomb interaction, (ii)
accurate calculation of phonon frequencies and electron-phonon matrix
elements,  and (ii) a well converged calculation of $\lambda$.
In any case, the modest values of $\lambda$ indicate weak coupling
and low but non-zero T$_c$ when augmented with $\mu^*$.
These results for two TRSB superconductors provide a clear indication 
that BCS pairing (spin singlet, phonon mediated, weak coupling) is
viable, and that other 
origins of the spontaneous magnetic field should be sought.

\subsection{$p$-wave versus $s$-wave}
The near (possibly complete) absence of anisotropic pairing states 
in conventional Fermi liquid
metals raises questions about why they seem to be so disfavored. After all,
the antisymmetric $p$-wave state should gain energy due to the two members
of the Cooper pair having a reduced short-range Coulomb repulsion.
Foulkes and Gy\"orffy\cite{Foulkes1977} looked at the scattering vertex 
function $\Gamma$, given
schematically by $\Gamma$=$I$-$IGG\Gamma$, where $I$ is the irreducible
scattering vertex and $G$ is the single particle Green's function. 
Assuming that $I$ does not depend on relative spin orientations, they
compared the coupling strength $\lambda_0$ for an $s$-wave kernel
to its $p$-wave counterpart $\lambda_1$. A key point is that for
$p$-wave, $\mu$ and therefore $\mu^*$ will be much reduced, perhaps to near 
negligible, so the
effective couplings $\lambda_0-\mu^*$ and $\lambda_1$ may be the
quantities that should be compared. 

Foulkes and Gy\"orffy took into consideration the accepted lore that
anisotropic pairs are strongly affected by impurity scattering, requiring
very clean metals. Reducing T$_{c,1}$ ($p$-wave pairing)
proportional to the transport
broadening $\hbar/\tau$, an accepted approximation, they suggested 
that $p$-wave pairing might occur in Pd, W, or Rh samples, but
requiring residual resistivity ratios between $10^3$ to $10^5$,
representative of {\it extremely} pure crystals. 
This work provided at least the plausibility of $p$-wave pairing
in conventional metals. Other works have suggested that phonon pairing
in conjunction with other interactions can generate odd-pairing for
certain choices of parameters.\cite{Schnell2006,Brydon2014}.

\subsection{Spin triplet pairing; INT model}
As mentioned above, the superconducting properties of this class of
fragile magnetic superconductors, barring the emergent magnetic
field, had been understood initially in terms of
singlet pairing, with results similar to related singlet superconductors
with low T$_c$. Nevertheless, triplet pairing has become the
favored direction of study for the order parameter in this TRSB class.
Ferromagnetic spin fluctuations have long been considered as the likely source of
triplet pairing,\cite{Berk1966} although triplet models based on
antiferromagnetic fluctuations have been constructed.\cite{Kreisel2022} 

In the formalism reviewed earlier, the symmetries have been triplet $S$
(even), $L$ symmetric and even (full space group symmetry), 
and the additional broken symmetry,
call it $G$, odd. The triplet $S$ space provides coupling to the magnetic field,
and receives the attention. The developers of the INT model,
Weng {\it et al.}\cite{Weng2016} and Ghosh {\it et al.},\cite{Ghosh2020}
posit this type of spin-triplet order parameter.
Their $G$ degree of freedom relied on breaking of band, Fermi 
surface, or orbital (near) symmetry,
which allowed the even triplet-spin symmetry to provide what was
necessary to account for a spontaneous magnetic field. This model has
evolved and merits an overview.

In their 2010 study of the symmetry restrictions in LaNiC$_2$, Quintanilla
{\it et al.}\cite{Quintanilla2010} arrived at an initial proposal
for the observation of TRSB. Specific properties they addressed were
(i) the non-centrosymmetric space group, (ii) the point group ($C_{2v}$)
with only non-degenerate irreducible representations, (iii) necessity
of considering spin-orbit coupling (SOC), (iv) the restriction that
the effect of SOC must be small, in spite of their observation that
SOC splitting of bands
around the Fermi level is more than an order of magnitude larger
than the SC gap. Satisfying these conditions, the only allowed order
parameter was distinctive: primarily singlet with a small admixture of triplet,
{\it i.e.} most Cooper pairs are (conventional) singlet, while a
few are equal-spin paired. There is however the necessity that
an additional symmetry of the normal state
if broken would lead to additional considerations.

Possibly concluding that the singlet+triplet state was unlikely,
Weng {\it et al.} constructed\cite{Weng2016}, and
Ghosh {\it et al.} refined\cite{Ghosh2020} an
Internally (antisymmetric) Nonunitary Triplet pairing (INT) spin-triplet
$S$=1 picture for LaNiC$_2$ and LaNiGa$_2$  in which there is
parallel S=1 spin pairing (the $S_z$=0 channel is a symmetric combination
of singlet pairs and would lie at higher energy).
For the magnetization,
$|$$\uparrow\uparrow> (S_z=+1)$ occupation exceeds (slightly) that of
$|$$\downarrow\downarrow> (S_z=-1)$. To prepare for discussion of a
singlet analog in Sec.~\ref{subsec:scenario} we note here that the
INT OP has the form
\bea
\Delta_k=\Delta_O~(i\sigma_z)~\vec d_k\cdot\vec\sigma~~(i\tau_z),
\eea
see Appendix~\ref{app:tripletOP} for further discussion.
Here $\vec\sigma$ is the spin Pauli vector matrix, $\vec\tau$
is the Pauli vector matrix in an auxiliary $r$ or $k$ space that provides
the symmetry to be broken simultaneously with $U(1)$ (superconductivity)
to satisfy the antisymmetry of the Cooper pair. 
The normalized $\vec d_k$ is a complex
vector relating to the direction of spin of the emergent spontaneous
magnetic field; possible $k$-dependence is not considered until
required by data.
This model is based on some electronic (quasi)degeneracy $\vec\tau$ --
degenerate bands or Fermi surfaces, or alternatively that of
an active atomic orbital on
symmetry-related atoms -- any might account for the necessary
additional symmetry to be broken. From some reported structure 
in the penetration depth
$\lambda_L(T)$ and specific heat $c_v(T)$ they argued
that a ``two gap'' (or ``two band'')
character might be responsible. Data for $c_v(T)$ obtained on single
crystals of LaNiGa$_2$ since these papers were published may also be
consistent with a single anisotropic gap.\cite{Badger2022}
Those interested in further implications of the INT model are 
referred to the original papers and to Appendix~\ref{app:tripletOP}, 
where a mathematically
similar model is proposed as an exotic singlet scenario.

\subsection{Challenges to triplet pairing}
Based on the assignment of triplet $S$=1 pairing in the class of materials
under discussion here, such exotic pairing was earlier considered as a
possibility only in correlated materials.
A more detailed understanding of the interaction and modern
interpretation, based on TRSB of weak coupling Fermi liquid metals,
 of triplet pairing appears
to confront several common expectations and challenges:\\
$\bullet$ that low T$_c$, weak coupling SCs commonly conform to Cooper's 
(BCS) singlet $S$=0, zero momentum pairing instability based on the
ever-present electron-phonon coupling, in terms of $\xi$, $\lambda$,
H$_{c1}$, and H$_{c2}$ that are understood on that basis\\
$\bullet$ triplet correlations will need to dominate singlet correlations
to break TRS, implying at least a moderately  strong triplet 
coupling (of unclear origin in an uncorrelated metal),\\
$\bullet$ dilute magnetic impurities affect (diminish) singlet (Cooper) pairing moderately; 
triplet states are predicted to be highly sensitive to defects,
even non-magnetic defects,\cite{Balatsky2006,Andersen2023} 
while in some of the listed  materials in Sec.~\ref{subsec:TRSBlist} T$_c$ 
values are consistent across sample quality,\\
$\bullet$ there are open questions relating to the superconducting 
state coexisting with
an intrinsic magnetic field;~\cite{Rosenstein2015} note that
the inferred spontaneous magnetic field is far weaker than
the observed (or derived) thermodynamic critical field $H_c$,\\
$\bullet$ only a select group of intermetallic compounds of 
seemingly similar non-magnetic Fermi liquid
character are spread across crystal classes,  across compositions, and 
for a variety of space groups with differing anisotropy, are identified 
as breaking TRS, and all are low T$_c$ materials,\\ 
$\bullet$ triplet pairing appears to require a wholesale reconstruction
of the pairing interaction, most of which (to repeat) are currently 
analyzed successfully in terms of singlet pairing.

This class of fragile
magnetic superconductors may be analogous to the situation 
for electron=-phonon coupling, where a small
uncertainty in a weak-coupling $\lambda$ can account for the difference
between low T$_c$ or none. A smaller than detectable spontaneous field will
lead to assignment to the conventional class, while a larger, just detectable
field, will lead to TRSB behavior. 
According to muon effects discussed here, all implanted muons should experience 
an emergent field and perhaps additional depolarization below T$_c$, 
but with some the field is smaller enough that it lies below detectability 
and thus is not included in the class of fragile magnetic superconductors.

Triplet $S$=1 pairing in this class has the sole, but crucial, 
justification that it justifies the observation of additional 
muon spin depolarization, or polar Kerr rotation, or (in one study) magnetization, 
below T$_c$. The conclusion has been that
the order parameter must couple to a magnetic field.
For triplet pairing, equal $S_z$=$\pm$ 1 occupations provide a degeneracy from
which symmetry can be broken by some small interaction.
As mentioned, other SC properties of this class inconsistent with 
triplet pairing. It is also a
challenge, for the future,  to identify a microscopic mechanism favoring equal 
spin pairing in this class, one
that overrides Cooper's strong favoring of both singlet pairing and symmetric zero
momentum pairs. These challenges are being addressed by theorists
material by material. Some generalities of triplet OPs, of which the INT 
model\cite{Weng2016,Ghosh2020} is one, are provided 
in Appendix~\ref{app:tripletOP}.

\subsection{Exotic singlet models}
\label{subsec:exotic}
Signals of TRSB have been proposed to be possible with singlet OPs, which
would however need to include some additional `exotic' components. One
approach was provided by Agterberg, Barzykin, and Gor'kov,\cite{Agterberg1999}
who considered a cubic material (later generalized) with isolated FS
pockets and competing attractive ($\lambda$) and repulsive ($\mu$)
interactions. A simple case built on three FS pockets, in which one of
the two OPs was a complex two-fold degenerate representation, and would
require TRSB to complete the Cooper pair antisymmetry. This  
section provides a brief description of one, and mentions a second,
different approach.

\subsubsection{An orbital magnetism scenario}
\label{sec:orbitalmoment}
Spontaneous valence orbital magnetism in crystalline solids is generally expected to 
be much smaller than conventional spin magnetization. However,
given the reported small field $\mu$SR values of 0.1-1 G, orbital magnetization would
only need to be the same as the spin magnetization discussed earlier, of the
order of 10$^{-3}-10^{-4}\mu_B$ per active unit in a cell.\cite{Ghosh2021d} 
Orbital currents have been 
discussed originally in the context of strongly correlated 
materials,\cite{Varma1997,Nayak2000}
and not yet verified. Independent searches in the layered cuprates 
have placed upper limits on the magnitude of such 
currents,\cite{Strassle2008} however
they remain of theoretical interest.

In such an event, the orbital current (on a [sub]nm radius scale)
would need to couple to the $U(1)$ order parameter (pairing).
As discussed in Sec.~\ref{subsec:supercurrent}, an orbital supercurrent is driven
by the muon moment that produces a local field, which in a 
susceptible system  might trigger a (local) orbital current state more broadly.

The orbital loop current picture of
Ghosh, Annett, and Quintanilla\cite{Ghosh2021d} provides a mechanism --
an OP satisfying the requirements -- in a specific model. Based on
the idea of a Josephson current flowing between equivalent atoms in
the primitive cell, they constructed 
a Ginzburg-Landau-like expression for the condensation energy 
involving right and left oriented current loops. The requirements were
that the OP satisfied the appropriate symmetries and that the free
energy was real. 

The phase diagram in terms of the free energy constants $\alpha$
and $\{\beta\}$ could be examined, finding regimes where the condensation
energy was positive (SC would occur). Their model carried their stated 
requirements for such orbital order parameter:
(i) there is on-site intra-orbital {\it singlet pairing}, 
(ii) more than two distinct but symmetry related sites within the unit 
cell are required to host the orbital current loop,
(iii) there must be a degenerate irrep in the point group, and
(iv) the free energy parameters require tuning. 
The only $k$-dependence arises from electronic form factors, and nodal
lines might occur across the Fermi surface but are not required by
symmetry.  
This model was proposed as a possibility for the compounds in the
Re$_6$(Ti,Zr,Hf) system. With its somewhat larger value of T$_c$ than 
others in this class, they explored for a maximum of the magnetization
and obtained a maximum field of the order of 1 G.
The stated criteria for this orbital current do not seem to correspond to 
most of the class of fragile magnetic superconductors,
including LaNiGa$_2$.

\begin{figure} 
\includegraphics[width=0.6\columnwidth]{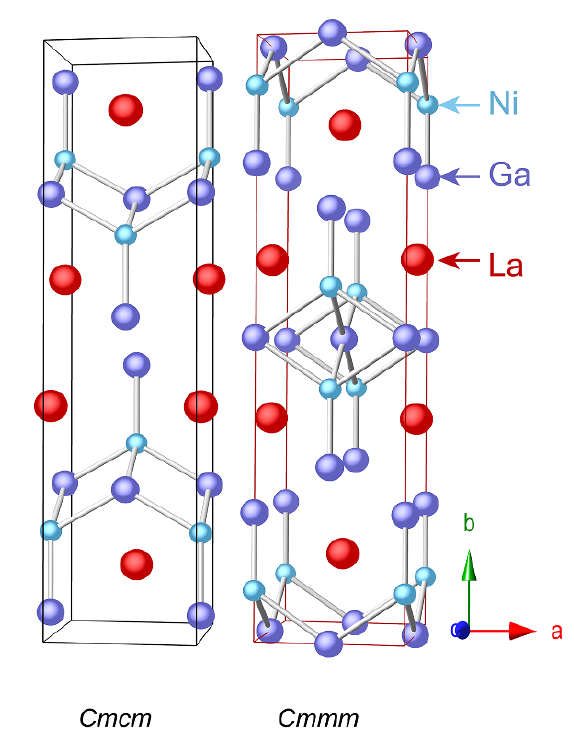}
\caption{Comparison of the single crystal $Cmcm$ structure of 
LaNiGa$_2$\cite{Quan2022,Staab2022}
compared with the $Cmmm$ structure previously obtained from
polycrystalline samples.\cite{Grin1982} The structural difference is mainly
in the Ni-Ga coordination. The electronic differences become
pronounced when specific features of the Fermi surfaces, viz.
pairs of Dirac points, affect physical properties, see text. 
Figure taken from Li {\it et al.}\cite{HLi2025}
} 
\label{fig:2structures} 
\end{figure}

\subsubsection{LaNiGa$_2$: spin singlet, Dirac point triplet}
\label{subsec:scenario}
\begin{figure}
 \includegraphics[width=0.65\columnwidth]{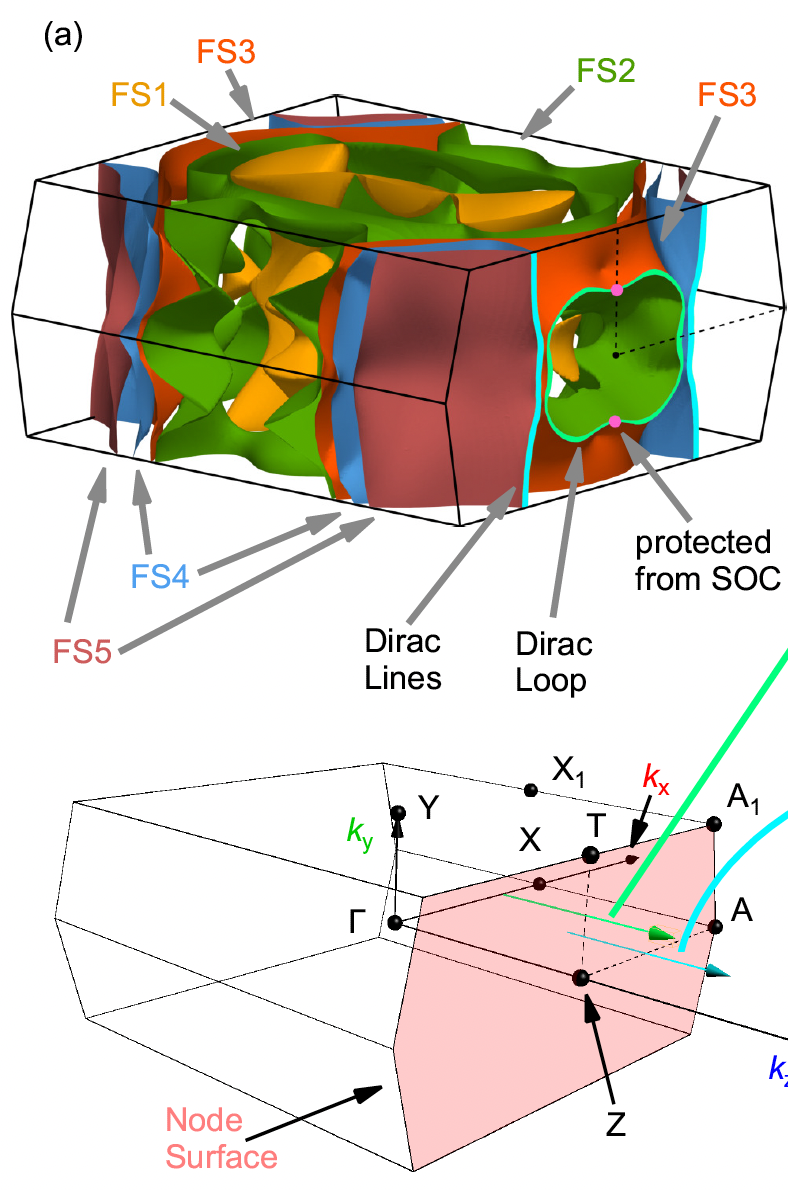}
\caption{Fermi surfaces and nodal loops of LaNiGa$_2$.
Top: five Fermi surfaces of $Cmcm$ LaNiGa$_2$, illustrating
the nodal line and nodal loop degeneracies on the zone surface,
which is pictured as the pink plane in the lower panel. 
The green and red Fermi surfaces (FS2 and FS3) ``stick together''
on the light green loop; the blue and brown Fermi surfaces
(FS4 and FS5) merge on the vertical blue
lines. All bands on the pink surface plane are four-fold degenerate
(spin+two bands) due to the non-symmorphic $Cmcm$ group symmetry.
Also: effect of spin-orbit coupling vanishes along the $Z$-$T$ line
emphasized in the lower panel.
The FS2/FS3 loop crosses the $Z$-$T$ line at the red dots,
denoted by ``protected from SOC,''
leading to 3D Dirac point character at those dots. 
Bottom panel: symmetry labels of $Cmcm$ Brillouin zone, node surface
shown in pink and the vertical $Z$-$T$ line shown.}
\label{fig:FS}
\end{figure}

{\it $Cmcm$ symmetry.}
A case study of LaNiGa$_2$ was chosen here because its Dirac points include 
it into the general class of topological superconductors, and it presents a
valley degeneracy that may be susceptible to breaking of symmetry
[time-reversal, inversion] pinned to the Fermi surface. The $Cmcm$ structure
is contrasted in Fig.~\ref{fig:2structures} with the earlier assignment 
($Cmmm$) from polycrystalline samples.
These Dirac points (DPs) were discovered by Quan {\it et al.}\cite{Quan2022} 
and confirmed by angle-resolved photoemission spectroscopy (ARPES) 
by Staab {\it et al.}\cite{Staab2022} and by a
de Haas-van Alphen study by Li and co-authors.\cite{HLi2025}.
These DPs are the remaining degeneracies after SOC splits the rest of a
degenerate loop of intersecting Fermi surfaces on the BZ face, as 
discussed in more detail in Appendix~\ref{sec:FS}. These
DPs lie on the intersection of two FSs -- independent of doping, strain, and other symmetry 
preserving variations -- if they cross the $Z$-$T$ zone-face symmetry line as 
shown in Fig.~\ref{fig:FS}, they are protected by $Cmcm$ crystal symmetry. 
This degenerate pair is closely analogous to the DPs in graphene,\cite{Neto2009}
-- a pair related by inversion and time-reversal, but in addition 
for LaNiGa$_2$ surviving SOC effects
and doing so in an orthorhombic 3D crystal. The non-analyticity of the 
electronic eigensystem provides a $N(E)$ non-analyticity
$\Delta N(\varepsilon)\propto |\varepsilon|^3/v^4$ at the Fermi energy 
($v$ is roughly the
geometric mean of the velocities along the principal axes) but in a
background metallic density of states $N(0)$.

Suppose that the Cooper pair entities couple to a standard (odd) zero momentum spin singlet,
which ties up the spin degrees of freedom. It remains to account for the spontaneous
magnetic moment below T$_c$ (assuming it does not arise from the supercurrent
as discussed in Sec.~\ref{subsec:supercurrent}). Each DP then has two orbital degrees of freedom,
which we denote $\nu_2$ and $\nu_3$, whose positions at $\vec k_D$ and $-\vec k_D$ are
pictured in Fig.~\ref{fig:FS}, labeled as {\it protected from SOC}. 
The second DP can also be pictured as at the mirror point along $Z$-$T$, which
is equivalent to $-\vec k_D$ by Brillouin zone translational symmetry.
Choosing an order for the DPs,
the four states ($\nu_2,\nu_2$), ($\nu_2,\nu_3$), ($\nu_3,\nu_2$), {$\nu_3,\nu_3$) comprise 
a 2$\times 2$ $\tau$ space for pairing, mathematically isomorphic to the spin-half
$\sigma$ space.  Without pair interactions, all four $\tau$ states are
degenerate. For definiteness, we define $\nu_2$ as the continuation of
band 2 as it approaches the DP, and similarity for $\nu_3$. Bands 2 and 3 are,
after all, two separate, orthogonal bands on two FSs, each with its own orbital 
(real space) Bloch state variation over the unit cell. 

{\it The order parameter.}
Considering a residual interaction beyond nonsymmorphic band theory (which 
{\it enforces} the degeneracy including SOC), these four pairings may have different energies --
residual interactions can break the $\nu_2,\nu_3$ symmetry and hence that of
the $\tau$ singlet and triplet combinations.  Note that in the normal state 
degeneracy implies a $\nu_2$$\leftrightarrow$$\nu_3$ symmetry (energy degeneracy). 
One should note that the two DPs are already correlated by the 
Cooper pair zero momentum $K_{cp}=0$ condition. 
The interest is in the even symmetry $\tau$ triplet states 
($\nu_2,\nu_2$), ($\nu_3,\nu_3$) ($\tau_z$=$\pm$1) to complete the Cooper pair antisymmetry
(keeping in the back of the mind some property that will activate a  spontaneous
magnetic field).  

The valley symmetry-breaking (VSB) OP takes the form 
(analogous to that of the INT form)
\begin{eqnarray}
\Delta = \Delta_o~(i\sigma_y)~\vec g_k\cdot\vec\tau~(i\tau_y),
\end{eqnarray}
with $\Delta_o$ specifying the amplitude and internal orbital pairing character 
provided by the normalized complex $\vec g_k$ vector that
denotes the `direction' in $\tau$-space (valley space) analogous to the $\vec d$
vector in spin triplet OPs, and $i\tau_y$ provides coupling to $U(1)$. At least 
in its simplest form, this OP implies simultaneous breaking of gauge, spin,
and valley symmetries, as well as the total momentum $K_{cp}$=0 favored
by Cooper pairs. 

This valley degeneracy thus provides the conceptual platform for VSB.
Dealing with trilayer twisted graphene with an analogous two-valley
symmetry, Gonzalez and Stauber considered the residual Coulomb interaction 
(beyond band theory) between carriers in the two valleys at the
Hartree-Fock level.\cite{Gonzalez2023} The energy splittings 
(symmetry breaking) in that case
depended on band filling and the degree of screening of the Coulomb
interaction between the two valleys, the result reflecting a VSB instability.
The main distinction in LaNiGa$_2$ is that the interaction between carriers 
in the two valleys will involve metallic screening and a conventional
electron-phonon coupling across the rest of the Fermi surfaces. At this conceptual
level we consider only the stoichiometric case.

{\it Emergent magnetism; spin-orbit coupling.}
The result of VSB obtained by Gonzalez and Stauber
is that the energies of the two valleys are split above and below the Fermi level,
which in the valley electronic structure of LaNiGa$_2$ ($E_F$=0) would be
$\varepsilon_{\pm\vec k_D}$ $=$ $\pm$ $\varepsilon_{vsb}$, implying also a bending
of the dispersion around each DP.  This symmetry breaking would be
a version of TRSB, but by loss of inversion rather than driven by a 
magnetic instability, a possibility that is not so widely considered. 
In the spinless case this would not involve magnetism. However, spin-orbit 
coupling is an integral feature in
the DPs in LaNiGa$_2$. Recall that apart from band anticrossing, SOC leaves
bands with ``pseudospin'' -- mostly up, or mostly down. 
With the degree of spin character in the two bands
varying with $k$ along the band, namely, through the DPs, a relative shifting of the DP
energies entailing a differing band filling, with unequal spin character 
above and below $E_F$, and would result in an imbalance of spin character. This 
mechanism results in magnetism as a parasitic result without required
magnetic instability.

{\it Gaplessness: unitarity on not.} 
This scenario would account for the singlet SC state properties\cite{Badger2022} of
LaNiGa$_2$, see Appendix~\ref{app:SCparameters}. 
With an appropriate choice of OP $|\vec g^* \times \vec g|$ $<$1,
one can model a fully gapped SC state as observed. It would be a
non-unitary, two BdG QP gap system (see Sec.~\ref{app:tripletOP}) 
as proposed for the INT model. The vector $\vec g$ would couple to an emergent VSB
field ${\vec{\cal G}}$ analogous to the magnetic field in the INT model.
The broken inversion symmetry gives TRSB ($\varepsilon_{dp1}$$\neq$
$\varepsilon_{dp2}$). Since bands 2 and 3 each has its own real space character, 
and given the normal state symmetry $\nu_2\leftrightarrow\nu_3$, it is natural 
that ($\nu_2,\nu_2$) and ($\nu_3,\nu_3$) coupling will be equal,
and possibly stronger than off-diagonal ($\nu_2,\nu_3$) pairing. 

While the scale of SOC band splitting, 30-40 meV in LaNiGa$_2$, is 
typically regarded as small, it is (i) the crucial feature in having
a pair of DPs rather than a loop of degeneracies, and (ii) it is large
compared to $2\Delta$, $k_B$T$_c$, the SC condensation energy, and the
energy associated with the small inferred magnetic field. Categorizing the possible
order parameters when (schematically) total angular momentum
${\cal J}=L+S$ defines the symmetries of the
normal state, rather than $L$ and $S$ separately, will be left for further
study. (Here $L$ characterizes the
symmetry in real space, or in the BZ, and SOC couples it to spin.)

{\it BdG quasiparticles.}
The Bogoliubov-de Gennes quasiparticles could have dispersions
analogous to those derived by Quan {\it et al.} for the spin case.\cite{Quan2022}
 The unitary and non-unitary possibilities are
 shown in Fig.~\ref{fig:QP-dispersion}, respectively gapped and gapless.
A difference from INT would be  
that its spin character would be interchanged
with the new valley character.
Although this scenario shares mathematical similarity to the INT model, it
is distinct physically, both in the type of symmetry breaking 
(and pairing interaction) and in
the importance of SOC in how the parasitic magnetization arises.
A full model would include the interactions between the various states.
This picture does have another similarity to the INT model: it does not transfer 
easily, if at all, to other members of the fragile magnetic superconductors.

\section{Sample character: defects, granularity, inhomogeneity}
\label{sec:sample}

\subsection{General cases}

It is not unusual for samples, and more particularly quantum materials, 
to display sample dependence of small signals, and 
reports of $\mu$SR depolarization are subject to sample type and quality,
see Sec.~\ref{subsec:TRSBlist}, and Sec.~\ref{subsec:YSR} for 
discussion of magnetic impurities.
In low symmetry, polycrystalline  samples, the measurement
will average over anisotropic responses, with the impact (loss of detailed
information) increasing with the degree of anisotropy of the crystalline
and electronic structure, and (thinking of $\mu$SR) lack of symmetry of 
the muon site. 

High Tc cuprates notoriously can be strongly inhomogeneous on the nm scale,
but this issue of sample irregularities arises more generally as atom-specific
probes are extended in capability.
Applying superconducting-tip scanning tunneling microscopy to a 
somewhat overdoped Bi-cuprate,
J. C. Davis has reported\cite{Davis2026} their imaging to reveal strong
inhomogeneity of the local SC energy gap such that there are 
Josephson-junction coupled SC regions (puddles).  The interfaces involve supercurrents
that form a spider-web of interconnected supercurrent rivers with
their associated magnetic fields, evidence of TRSB at the 10 nm scale
and characteristic of granularity of the superconducting order parameter.

Andersen, Kreisel, and Hirschfeld have considered possible mechanisms behind 
spontaneous TRSB due to disorder in SCs,\cite{Andersen2023} 
which includes cases of chiral OPs less closely related to the present discussion.
They note that non-magnetic defects may -- in quantum materials ({\it i.e.}
close to magnetic order) -- give rise to magnetic centers that can
affect signals from several probes. Magnetic impurities give rise to 
YSR bound states as discussed here; that the muon is a magnetic impurity
has been at the center of the present discussion. A muon may be attracted
to regions of defects, and even neglecting this, $\mu^+$ is typically found to
lie in an interstitial position of low symmetry. Two specific classes of materials 
that have been studied by $\mu$SR and illustrate sample dependence follow.

\subsection{U-based heavy fermion SCs}

Sample type (polycrystal or single crystal) and quality have led to
differing conclusions in several cases. Uranium-based SCs have provided
a few examples. 

$\bullet$ UPt$_3$.
The hexagonal heavy fermion SC UPt$_3$ displays SC
transitions at 0.50 K and 0.45 K.
An early $\mu$SR report found TRSB only below 0.45 K, with a quoted 
field of 0.1G.\cite{Luke1993,Luke1993a} 
A  PKE signal, also in the low T$_c$
phase of a crystal,\cite{Schemm2014} did not specify a field value. 
Following work on single crystals found\cite{Dalmas1995} no $\mu$SR
signal below T$_c$. Further study\cite{Higemoto2000} extended $\mu$SR study of a single
crystal cylinder down to 21 mK, without indication of TRSB.

$\bullet$ In U-based heavy
fermion SC U$_{1-x}$Th$_x$Be$_{13}$, substituting with Th introduces
disorder in addition to carrier doping. Reported from $\mu$SR
was an increased depolarization\cite{Heffner1990} 
with onset below 0.4 K for 
$x$=0.019-0.035 (other experiments had indicated two SC
transitions, the upper one sometimes around T$_c$=0.55-0.65 K).
A later neutron scattering study\cite{Hiess2002}
found no evidence of a field in an $x$=0.035 sample.
Further references on heavy fermion SCs can be found in
the review of Stewart,\cite{Stewart2019} which seems to leave the issue
for this system undecided, while the overview of Ghosh 
{\it et al.}\cite{Ghosh2021} addresses this system for $x$=0.02-0.04 as TRSB.

$\bullet$ UTe$_2$. 
After the discovery of unusual phases of UTe$_2$ (in temperature, field,
and pressure), there were reports
(2019-2023) supporting TRSB from three NMR 
studies\cite{Nakamine2019,Fuji2022,Matsumura2023} and two reports of a 
nonzero PKE signal.\cite{Hayes2021,Wei2022} Sample type
and quality was reported and discussed in these papers, noting
that positive or neutral signals could be observed in various
regions of single crystals.
$\mu$SR measurements on molten flux-grown single crystal samples 
in 2023 by Azari {\it et al.}\cite{Azari2023} reported no observable $\mu$SR
signal.  Sample quality, possibly involving
U vacancy complexes that could become magnetic, was suggested to be
a source of the earlier experimental indications.
Further PKE studies\cite{Wei2022,Ajeesh2023} found no evidence of a TRSB signal.

\vskip 2mm
Considering other correlated systems, measurements on two cuprates,
where there have been theoretical suggestions\cite{Varma1997}
of orbital loop derived magnetic fields, have put upper
limits on such a field, which has not been detected. For
the cuprate high temperature SCs (80-93 K) YBa$_2$Cu$_3$O$_7$ and
Bi$_2$Sr$_2$CaCu$_2$O$_8$, no additional depolarization was detected
below T$_c$.\cite{Kiefl1990} 

\subsection{The unusual case of Sr$_2$RuO$_4$}
The enhanced Fermi liquid SC Sr$_2$RuO$_4$ (T$_c$=1.5K) exhibited (1998) a
$\mu$SR signal\cite{Luke1998}, followed by a 2009 PKE study\cite{Kapitulnik2009} 
supporting TRSB (signal near the limit of resolution).
A 2003 review\cite{Mackenzie2003} of considerable
experimental data on Sr$_2$RuO$_4$ and potential order parameters
concluded that triplet pairing was a likely possibility. Tsuruta
{\it et al.} around 2015 presented a model\cite{Miyake,Tsuruta2015,Miyake2017} 
for an impurity in a chiral
superconductor (as Sr$_2$RuO$_4$ had been suggested to be) that, when
applied to $\mu$SR suggested an induced field of roughly the size
that had been reported.
A 2019 overview presented accumulating evidence supporting triplet 
pairing;\cite{Wysokinski2019} in 2021 a $\mu$SR signal below T$_c$ was found
under pressure in Sr$_2$RuO$_4$ single crystals.\cite{Grinenko2021a} 
A related group reported supporting results under strain.\cite{Grinenko2021b}  
A theory\cite{Miyake} supporting this was built on an energy-dependent density 
of states, viz. near a van Hove singularity, and using parameters from 
Sr$_2$RuO$_4$ predicted a spontaneous TRSB field of the order $10^{-2}$ G,
not detectable by probes currently in use. 

During this time period proposals of singlet
pairing for Sr$_2$RuO$_4$ with exotic real space order parameters 
appeared\cite{Kivelson2020,Romer2020,Benhabib2021,Grinenko2021,Leggett2021} 
as new samples\cite{Maeno2024a} and more data accumulated.
A 2024 review emphasized that the issue of TRSB in Sr$_2$RuO$_4$ was not
settled,\cite{Mackenzie2023} while a  2024 survey\cite{Maeno2024b} stated 
``...a new development overturned past experimental results, and 
spin-singlet-like behavior became conclusive,'' followed by an agreeing 
comment referring to reversal of consensus in the field.\cite{Mazin2022}
Singlet pairing in this compound had been 
entertained earlier\cite{Schnell2006}
as measurements do not exclude a field of orbital origin, although models
tended to require tuning of parameters or an exotic form of real-space
character, viz. chiral $d$-wave. A relevant overview referenced more
than twenty suggestions\cite{Holmvall2023} of chiral $d$-wave order parameters
that will host TRSB, including some in the list in Sec.~\ref{subsec:TRSBlist}. 
This precession of numerous experiments and evolving evaluations reflects how delicate
the determination of TRSB and exotic order parameters has been, with sample 
variation playing a large role.

\section{Summary}
\label{sec:summary}
This paper has focused on a class of normal Fermi liquid, low T$_c$,
metals that have been identified (primarily) by $\mu$SR depolarization data
as time-reversal symmetry breaking SCs, in the midst of a few others giving
negative results. This class has been denoted fragile
magnetic superconductors, because the inferred
magnetization is no more than 10$^{-3}\mu_B$ per atom, near the limit
of modern methods.

A few points can summarize the discussion, in some rough order of importance.\\
$\bullet$ Superconducting properties -- critical fields, coherence lengths,
penetration depth, TRSB signals, other data --should be considered
{\it in toto}\cite{Mazin2022} in contemplating the relevant order parameter. 
All SCs are initially analyzed as singlet pairing, and the weakly coupled SCs discussed
here have typical singlet SC properties. Unconventional SCs are those that
do not conform, and also have unconventional (non-Fermi liquid) normal states.\\
$\bullet$ In $\mu$SR experiments, time reversal symmetry is broken, strictly speaking, when
the muon is implanted and coupled with the solid, with magnetic effect felt most 
strongly near the muon site where the measurement (muon decay) originates.\\ 
$\bullet$ Conventional theory, giving a circulating supercurrent around the
muon proportional to the muon vector potential, will give rise to a
magnetic field in the region of the muon in any superconductor.
Quantum theory of the short-range supercurrent, hence the magnitude of the
field at the muon site, remains to be elucidated.\\
$\bullet$ The spontaneous field, detected primarily by zero-field $\mu$SR studies,
appears at T$_c$ (occasionally below or above T$_c$), with a value
near the limit of detection, and one that 
seems challenging to estimate theoretically, or to support with known mechanisms.\\ 
$\bullet$ With TRSB signals near the lower limit of detection, it is
normal that some superconductors with weaker field at the muon
will give no detectable signal and thus be classified as 
conventional, while those with somewhat stronger
fields that extend above the detectability limit 
are designated as TRSB. \\
$\bullet$ Several systems have shown differences due to sample
quality, with some signals disappearing in higher quality
single crystal samples while other properties are little affected.\\
$\bullet$ Calculated T$_c$ of singlet, phonon-pairing SC, 
in two of the fragile magnetic superconductors, LaNiC$_2$ and
LaNiGa$_2$, indicate that phonon coupling is sufficient to
account for singlet pairing and low T$_c$.

\section{Acknowledgments}
I acknowledge important continuing interactions with V. Taufour, 
N. J. Curro, I. Vishik, H. Hingorani, and R. R. P. Singh,
and helpful references to the literature from 
J. Autschbach, D. M. Ceperley, and J. Terning.
Early parts of this work were supported by National Science
Foundation Grant No. DMR 1607139.

The author states no conflicting interests.



\section{Appendices}

\subsection{Induced polarization}

\subsubsection{{Field at $\mu^+$ site from magnetic polarization}}
\label{app:inducedpolarization}
{\it Classical treatment.}
Magnetic polarization to an applied (``external'') field is usually
treated in the quasiclassical approximation of spin up versus spin down.
However, each electron has its own vector potential arising from its
Bohr magneton magnetic moment, which in the electron gas is commonly
averaged out. The effect can be modeled as follows.
Each volume element of electron moment
$\vec{M}^{ind}(\vec{r})\Delta V$ will produce from its magnetic moment
the same form of dipole field $\vec{B}^{\mu}(\vec r')\Delta V$
from the magnetization from $\vec r'$ via $\chi_p \vec B^{\mu}(\vec r)$ 
 as given by the dipole
expression, except that the original origin $\vec{0}$ will be
assumed by $\vec{r}$ and the position of a given field point will be
$\vec{r'}$.

Then
\begin{eqnarray}
\vec{\mathcal{B}}(\vec r')&=&
   \int d^3r \frac {3(\widehat{r'-r}) \vec{M}^{ind}(\vec{r})\cdot
     (\widehat{r'-r}) - \vec{M}^{ind}(\vec{r}'-\vec{r})}  {|r'-r|^3}.\nonumber
\label{eqn:integral0}
\end{eqnarray}

Simplification occurs
because we are only interested in the field at the muon site,
{\it i.e.} at $\vec{r'} \rightarrow 0$, so $\widehat{\vec{r'}-\vec{r}}
\rightarrow -\hat{r}$. Note: $\hat{z}=(0,0,1)$ is a
direction and remains unchanged. Then by change of coordinates
\begin{eqnarray}
\vec{\mathcal{B}}(0)&=& \int d^3r \frac {3(-\hat{r}) \vec{M}^{ind}(\vec{r})\cdot
     (-\hat{r}) - \vec{M}^{ind}(\vec{r})}  {r^3} \nonumber \\
 &=& \int d^3r \frac{3\hat{r} \vec{M}^{ind}(\vec{r})\cdot
     \hat{r} - \vec{M}^{ind}(\vec{r})}  {r^3}.
\label{eqn:integral}
\end{eqnarray}


By cylindrical symmetry (of the assumed HEG) only need the $z$-component 
of the field is non-zero.
The dot product in Eq.~\ref{eqn:integral} 
\begin{eqnarray}
\vec{M}^{ind}(\vec{r})  \cdot  \hat{r}
   &=&\chi_p(r)  \vec{B}_{\mu}(\vec{r})  \cdot \hat{r} \nonumber \\
 &=&\chi_p(r)  \frac{\mu}{r^3}
         (3\frac{xz}{r^2}, 3\frac{yz}{r^2}, 3\frac{z^2}{r^2}-1)
          \cdot (x,y,z)/r \nonumber \\
 &=&\chi_p(r)   \frac{\mu}{r^3}
      \left[   ( 3\frac{x^2}{r^2} + 3\frac{y^2}{r^2})
             + (3\frac{z^2}{r^2}-1)\right] \frac{z}{r} \nonumber \\
   &=&\chi_p(r)  \frac{2\mu}{r^3}\frac{z}{r}
     =\chi_p(r) \frac{2\mu\cos\theta}{r^3}.\nonumber \\
 3\frac{z}{r}[\vec M^{ind}(\vec r)\cdot \hat{r}]
   &=&   \frac{\mu}{r^3} \chi_p(r)=\mu\frac{6z^2}{r^2}.
\end{eqnarray}
This result is non-negative, positive along the $\hat{z}$-axis 
and vanishing in the $x$-$y$ plane. It is worthy of note that the 
$x$ and $y$ dependence has dropped out of the dot product of
two vectors depending (apparently) independently on $\vec r$,
again an effect of symmetry.

After subtracting off the $M^{ind}_z(\vec r)$ term in the dipolar
field that is large along the $\hat{z}$-axis, negative in the
$x$-$y$ plane, the $z$-component of $\vec B(0)$ involves some 
cancellation: the `incoming' dipolar field from the muon is
interfered with by the `outgoing' dipolar fields from each point
$\vec r$. The result becomes 
\begin{eqnarray}
B_z(0)&=&\int d^3r \chi_p(r)
  \left[  \frac{3z}{r}~(\frac{2\mu}{r^3}\frac{z}{r})
     - \frac{\mu}{r^3}(3\frac{z^2}{r^2}-1) \right] / {r^3} \nonumber \\
  &=& \mu \int d^3r \chi_p(r)
  \left[  6 \frac{z^2}{r^2} - (3\frac{z^2}{r^2}-1) \right] / {r^6} \nonumber \\
 &=& \mu \int r^2~dr~d\nu~d\phi \chi_p(r)
  \left[ 3\frac{z^2}{r^2} + 1)  \right] / {r^6} \nonumber \\
  & =& 8\pi \mu~\int \chi_p(r) \frac{dr}{r^4},
\end{eqnarray}
leaving, at this level of discussion, a divergent field at the muon site.
The strong small-$r$ divergence is startling, but
at small $r$ non-linear response (substituting for $\chi_p$) and quantum
effects require reconsideration.
Discussion of the resolution of this unphysical result is given in 
Sec.~\ref{sec:QuantumEffects}.

{\it Operator treatment.} The magnetic field of the muon, with magnetization 
 $\vec M^{\mu}(\vec r)=\vec{\mu}_{\mu}\delta(\vec r)$, 
gives rise to the well known dipolar magnetic field.
The linear relationship suggests the definition of an integral  ``dipolar operator'' 
${\cal D}(\vec r - \vec{r'})$ by 
\begin{eqnarray}
\vec {\bar B}(\vec r)&=&\int d\vec{r'} 
     {\cal D}(\vec r,\vec{r'})\vec{\bar{M}}(\vec{r'})\nonumber \\
 &=& \int d\vec{r'} \frac{3~\widehat{\vec r-\vec{r'}}
 [ \widehat{\vec r-\vec{r'}}\cdot \vec{{\bar M}}(\vec{r'})]-\vec{{\bar M}}(\vec{r'})}
        {|\vec r-\vec{r'}|^3},
\end{eqnarray}
which gives the magnetic field $\vec{\bar B}(\vec r')$ due to any magnetization field 
 $\vec{\bar{M}}(\vec r)$.
Applying this to the $z$-oriented $\mu^+$ point magnetic moment gives
Eq.~\ref{eqn:dipoleH} and is visualized in Fig.~\ref{fig:dipole}.  

The field
\begin{eqnarray}
\vec B^{\mu}(\vec r)=\int d\vec{r'} 
     {\cal D}(\vec r,\vec{r'})\vec{\mu}\delta(\vec{r'});
   ~~~\vec B^{\mu} = {\cal D}\vec M^{\mu},
\end{eqnarray}
polarizes the electron density according to the susceptibility $\chi_p$,
$\vec M^{ind}(\vec r)=\chi_p(n(\vec r)) \vec B_{\mu}(\vec r)$~
[in operator notation $M^{ind}=\chi_p B^{\mu}$]. 
This electronic magnetization (excess of $\uparrow$ moments over $\downarrow$)
gives rise to its own magnetic field which
becomes, in terms of the dipole operator
\begin{eqnarray}
\vec B^{ind} &=& {\cal D} \vec M^{ind}={\cal D}\chi_p \vec B^{\mu} 
       = {\cal D} \chi_p {\cal D}\vec M^{\mu}; \nonumber \\
\vec{B}^{ind}(\vec r)&=&\int d^3r' \int d^3 r'' {\cal D}(\vec r,\vec r') \nonumber \\
  & &~~\times~~\chi_p(n(\vec r')){\cal D}(\vec r',\vec r'')\vec M^{\mu}(\vec r'')
  \nonumber \\
\vec B^{ind}&=& {\cal D} \chi_p {\cal D} \vec M^{\mu}.
\end{eqnarray}
There are two important simplifications. First, for connecting to $\mu$SR
data we want only the on-site $\vec r=0$
field. Second, the original field arose from the point muon $\vec {r'}=0$, which
removes the second integral. In addition, $n(\vec r)$ varies smoothly by a factor
of two or less in an interstitial site in a metallic compound, and 
$\chi_p(n)$ is a regular and moderately varying function of $n$,\cite{Janak1977}
so we pull a representative value ${\bar{\chi}}_p$ out of the integral. This
leaves
\begin{eqnarray}
\vec{B}^{ind}(0)\approx {\bar{\chi}}_p \int d^3r' {\cal D}(0,\vec {r'})
     {\cal D}(\vec {r'},0)\vec{\mu}.
\end{eqnarray}   
This integral has the look of ${\bar{\chi}}_p<{\cal D}^2>\vec{\mu}$, and
indeed ${\cal D}(-\vec r)={\cal D}(\vec r)$ bears out this positive integrand.

The magnetization $M^{ind}(\vec r)$ operated on by the dipolar
operator ${\cal D}(0,\vec {r'})$ gives the induced magnetic field. 
Simplifying the integration variable to 
cylindrical coordinates with $z$-component of ${\hat r}=z/r)$:
\begin{eqnarray}
\vec{B}^{ind}(0)&=&{\bar\chi}_p \int d^3r \frac{1}{r^3}[3{\hat r}\cdot
  [\frac{3{\hat r}({\hat r}\cdot(\vec{\mu})-\vec{\mu}]}{r^3}] \nonumber \\
      & &   -[\frac{ 3{\hat r}({\hat r}\cdot \vec{\mu})-\vec{\mu}}{r^3}]\nonumber \\
 &=&{\mu \bar{\chi}}_p  \int \frac{d^3r}{r^6}
   3{\hat r} \Large[ 3\frac{z}{r} -\frac{z}{r}\Large]
  - \Large[3{\hat r}\frac{z}{r}-1\Large].
\end{eqnarray}
The $z$-component is
\begin{eqnarray}
B^{ind}_z(0)&=&\mu \bar{\chi}_p \int \frac{d^3r}{r^6}
   [9\frac{z^2}{r^2}-3\frac{z^2}{r^2}]-[3\frac{z^2}{r^2}-1]\nonumber \\
  &=&\mu {\bar\chi}_p \int_0^{\infty} \frac{2\pi~r^2dr}{r^6} 
     \int_{-1}^{+1}d\nu~(3\nu^2+1),
\label{eqn:diverge}
\end{eqnarray}
using cylindrical coordinates ($\nu$=$cos\theta$). 
This procedure reproduces the more direct result  
of Appendix~\ref{app:inducedpolarization}, a divergent integral that must be regularized
by quantum or relativistic near-field effects. 
All of this is complicated by the zero point uncertainty of the muon
position, discussed in the next section.

\subsection{Supercurrent: three theories}
\label{app:3theories}
There has been a progression of theory of the supercurrent, each providing
a linear relation between $\vec J^s$ and an applied vector potential $\vec A$,
necessary to account for the Meissner effect. They 
differ substantially following progress in condensed matter theory.
Only the final expressions will be given here, the
first two are textbook material, the third follows from BCS
theory. Each assumes a properly gauged vector potential.
A somewhat more expanded discussion is provided in
Ref.~[\onlinecite{superfluid}].

{\it London theory.}
For an inhomogeneous magnetic field, the London brothers obtained
the relation\cite{London1935}
\bea
\vec J^s(\vec r,T)
    & =& - \frac{4\pi}{\lambda_L^2(T)} \vec A(\vec r), \\
 \frac{c^2}{\lambda_L^2} &=& 4\pi e^2 \frac{n_s}{m},
\label{eqn:Londoncurrent}
\eea
where a superelectron density $n_s$ and mass $m$ were unspecified properties
(and were later put into electronic structure expressions\cite{London1948}).
This is a simple, local and temperature dependent, proportionality.
This expression arises from (in
addition to characteristics of a SC) screening by a
charged plasma, with material-dependent London penetration
depth $\lambda_L(T)$. Supposing $m$ is of the order of the free electron value, 
and choosing Nb as an example -- it is a borderline Type I - Type II SC,
with bcc lattice constant of 3.3\AA~and the accepted penetration depth of
$\lambda^{Nb}\approx 47 nm$ -- its London superfluid density is 
$\lambda_L^{Nb}\sim 5\times 10^{-7}$ electron/atom. Type II SCs have penetration
depths 1-2 orders of magnitude larger, and $n_s\propto \lambda^{-2}$ will give
even smaller values of superfluid densities (see Ref.~(\onlinecite{superfluid})
for further investigation of this question).  Nevertheless, the London expression 
remains in the minds and the publications of many researchers.

{\it Ginzburg-Landau theory.} Following from a Landau free energy
functional involving a complex order parameter $\Psi$, (not
to be confused with a manybody wavefunction) later
established to be proportional to the SC gap. The 
Ginzburg-Landau result\cite{Ginzburg1950}  was
\begin{eqnarray}
\vec J^s_{GL}(\vec r,T) &=& \frac {-ie^*\hbar}{2m^*} 
           Re[\Psi^*(\vec r,T)\nabla\Psi(\vec r,T)]
 \nonumber \\
&  &  -\frac {(e^*)^2}{m^* c} |\Psi(\vec r,T)|^2 \vec A(\vec r).
\end{eqnarray}
Later developments identified $e^*=2e$ as the charge of a Cooper pair,
$|\Psi|^2$ related to the density of Cooper pairs, and $m^*$ being
the Cooper pair mass.
The last term is a London-like expression of a material dependent
quantity times $\vec A(\vec r)$, however this generalized ``penetration
depth'' is position as well as temperature dependent. There is in
addition a new term involving the gradient of the order parameter,
which easily simplifies\cite{Ketterson1999} to the gradient of 
the phase of $\Psi$. This equation must be solved self-consistently
with one for $\vec A$ and typically involves boundary conditions.
Within Ginzburg-Landau theory,
Ashcroft and Krusch found\cite{Ashcroft2020} that for
a $\delta$-function magnetic impurity in a Type II SC and within a range of model
parameters, a localized {\it magnetic} impurity behaves similarly to
a magnetic vortex, the two being related by a
gauge transformation in their model.

{\it BCS theory.} This revolutionary theory established that SC is a low
energy property of a fermionic system, with transport and
thermodynamic phenomena involving only dynamically available 
carriers within $\sim\pm k_BT$ of the Fermi surface in the normal
state, or within a few times the gap $2\Delta$ in the SC state. F. London 
had earlier discarded his phenomenological parameters $n_s$ and $m$,
instead incorporating the language of Fermi surfaces and velocities into
the theory of the penetration depth.\cite{London1948} 
Chandrasekhar and Einzel\cite{Chandra1993} derived 
the corresponding relation from BCS theory,\cite{BCS1957} obtaining 
a tensor with diagonal terms for orthorhombic and higher symmetry structures
($j$ is a Cartesian index)
\bea
\vec J_{BCS,j}^s(\vec r,T)&=&-\frac{e^2}{c} \sum_k
                            \big[-\frac{\partial n_k}{\partial\varepsilon_k}
                        + \frac{\partial f(E_k)}{\partial E_k}\big]
                      \vec v_{k,j}^2 \vec A_j(\vec r) \nonumber \\
  &\rightarrow& -\frac{e^2}{c}N(0)v^2_{F,j} \nonumber \\
   &\times&  \Bigl[1-2\int_{\Delta}^{\infty}[-\frac{\partial f(E)}{\partial E}]
              \frac{E~dE}{(E^2-\Delta^2)^{1/2}}\Bigr]\vec A_j,   \nonumber \\
 \vec J_{BCS}(\vec r,T)
           &=& -\frac{4\pi}{\lambda^2_{BCS}(T)}\vec A_j(\vec r).
\label{eqn:BCSsupercurrent}
\eea
where the last expressions are for an isotropic gap $\Delta(T)$ 
(obtained from BCS theory) and
applies to orthorhombic or higher symmetry crystals.
In this expression $E_k=\sqrt{\varepsilon_k^2+\Delta(T)^2}$ 
is the energy of an excited quasiparticle
in state $k$, $\varepsilon_k$ is the normal state electron energy,
$f(E)$ is the Fermi distribution,
and $n_k$ is the single particle state occupancy. This expression gives
a substantially different picture of the response to $\vec A$: 
the first term is a T-{\it independent}
diamagnetic current similar in form to the London counterpart, and
the second term is a paramagnetic current from the excited quasiparticles,
which is zero at $T$=$0$ and grows to cancel the diamagnetic current
at $T_c$. 

This expression also differs by involving definitions of the
constants in terms of materials properties: the Fermi level density 
of states $N(0)$ (the measure of available dynamic electrons) and the
Fermi velocity $v_F$ (the speed of response). 
The T-dependence arises from that of the gap
$2\Delta(T)$, also given by BCS theory. This BCS expression holds for the
uniform bulk, thus with no $\vec r$ dependence. Usadel 
theory\cite{Usadel,Ketterson1999} based
on averaged Gor'kov Green's functions provides a generalization of SC
theory to cases
where the gap $2\Delta(\vec r,T)$ is position dependent due
to disorder, fields, or boundaries.
A general introspection into supercurrents in SCs was
provided by Koizumi and Ishikawa.\cite{Koizumi2020} While the aspect of 
a background degenerate fermion system was not taken into account,
this work gives vivid impression of the complexity of the 
supercurrent problem, and should be useful for further considerations.

\subsection{Kondo impurity in an exotic superconductor}
\label{app:Kondo-exotic}
This appendix follows from Sec.~\ref{subsec:YSR}, providing an example of a
Kondo impurity within an exotic order parameter state in a spherical model.
The question of impurity-induced magnetic fields in unconventional
SCs was addressed by Choi and Muzikar,\cite{Choi1989} taking an OP
of the $^3He$ $A$-phase as an example: $\Delta(\hat{k})=
(\hat{k}_x+i\hat{k}_y)
\Delta_o$, with orbital angular momentum character of +$\hbar$. 
This order parameter describes an internal
orbital momentum of the pair characterized by symmetry as
`ferromagnetic,' and the
resulting supercurrents produce a magnetic field at the impurity site.
Taking parameters thought at that early time to be characteristic of
YBa$_2$Cu$_3$O$_7$, but with sizable uncertainty, they estimated
the emergent field. 

The Choi-Muzikar approximation in general form for the on-site 
impurity magnetic field was
\bea
B(0)=-\hat{z}\beta(T)(ek_F^2)(\frac{T_c}{T_F})^2\frac{v_F}{c}
    (\sigma k_F^2).
\eea
This result is expressed in nearly-free electron language, hence
involving generic parameters.
Here $\beta(T)$ is a numerical factor that varies between zero and
unity, $ek_F^2$ is a field value expected to be of order
$10^2$~T,\cite{Choi1989}
$T_F$ is the Fermi temperature of the order of $5\times 10^4$K
for a large multisheeted Fermi surface metal ($E_F$$\sim$$4-5$ eV),
$v_F$$\sim$$c/300$ for good metals, and $\sigma$
is the impurity cross-section for which $\sigma k_F^2$ is
expected to be of the
order of unity. For this set of parameters they estimated 
the field as roughly $10^{-3}$G. 

\subsubsection{{YSR states in real materials}}
\label{app:realYSRstates}
Computational developments in treating defects (large unit cells in DFT)
and in modeling the Bogoliubov-de Gennes (BdG) band structure of real
superconductors has substantially
extended the understanding of the spectra and energetics of YSR states
in actual materials. Extending scanning tunneling methods to the study
of magnetic atoms or molecules on superconductor surfaces has provided
direct evidence of the behavior caused by YSR states.\cite{Kuster2021}   

The extension of the multiple scattering method (KKR: Korringa, Kohn,
Rostoker) to handle very large unit cells opened up the study of truly
isolated impurities. Similar extensions of band theory to the 
superconducting state -- the gap, BdG quasiparticle
spectrum -- for actual materials has played an important role in
furthering understanding.\cite{Saunderson2020} These
methods have enabled the study of YSR states, with real atomic moments
in (or on the surface of) materials being studied in labs. The chosen SCs include
the second discovered SC (Pb, T$_c$=7K) and the best elemental SC (Nb, T$_c$=9.2K) and
a possibly unconventional SC, NbSe. Fe, with its large and very well studied
moment, provides the preferred magnetic atom. Rare earth atoms, 
with their larger but more local moments, also invite 
attention. At present, applications incorporate phenomenological attractive
pair potentials to simulate the SC state of the host.

DFT methods and applications to the nominally non-magnetic N impurity in 
Nb have been described by Saunderson {\it et al.},~\cite{Saunderson2020}
who studied the effect on the gap and the excitation spectrum above the gap. 
Park {\it et al.}~\cite{Park2021} studied the spectrum, including a
zero bias peak, for a magnetic impurity (Mn, Fe, Co) in the $s$-wave SC
Pb (T$_c$=7K). 
The atomic $3d$ series embedded in Pb was studied by Ruessmann {\it et al.}
\cite{Ruessmann2023}, who obtained strong magnetic-SC coupling and orbital
splitting of a number of gap states (or resonances), sometimes extending 
across most of the gap. Spin-orbit coupling was shown to have a strong
effect on the spectrum, as spin-orbit splittings are much larger than the gap
of $\sim$2.5 meV. For the Fe impurity the gap is essentially closed,
{\it i.e.} SC order is effectively quenched in the vicinity of the impurity.
Topological SC was brought into study by Chiu and Wang,\cite{Chiu2021}
who analyzed the system Fe on Fe(Te,Se) with their topological $Z_2$ bands.
Their study indicated that the topological character of the bulk becomes
reflected in new behavior of the surface YSR states. 

Several groups have reported spectroscopic studies, including YSR states,
on SC surfaces decorated with magnetic entities.
\cite{Choi2017,Oppen2021,Xia2022,Saunderson2022}
The orbital structure of YSR states resulting from Cr atoms on SC Pb(111) 
was reported by Choi {\it et al.}\cite{Choi2017}, who found that the
Cr-derived YSR resonances extended across much of the gap but left a
pseudogap of low DOS around mid-gap ($\Delta=2.7$ meV).  
Xia {\it et al.}\cite{Xia2022} studied YSR states of a Kondo molecular 
magnet Tb$_2$Pc$_3$ layer on the Pb(111) surface.

\section{A case study: {\sc LaNiGa$_2$}}}
\label{sec:casestudy}

{\it TRSB in LaNiC$_2$ and LaNiGa$_2$.}
$\mu$SR observation of a depolarization increase near and below
T$_c$ on a polycrystalline sample was
reported in 2009 for LaNiGa$_2$,
with a field value later provided\cite{Ghosh2021} as 0.1 G.
A 2015 report\cite{Sumiyama2015} from magnetization
measurement detected a field at the 0.01 mG
($\mu$T) level, and only for $c$-axis orientation.
Their observed signal onset occurred {\it above} 
T$_c$=2.7K, in the 2.7-3.0K temperature region.  
A 2026 NMR study found a robust Hebel-Slichter peak in the
spin-lattice relaxation 1/$T_1 T$ below $T_c$; this signal
is understood to be one of the most definitive indications
of singlet pairing. 
Regarding LaNiC$_2$, Sumiyama {\it et al.} noted
that inferred fields are sample dependent. A $\mu$SR 
2021 study\cite{Sundar2021} on a single crystal
showed weak depolarization, and well below T$_c$. The
$T$-dependence suggested a two-gap system. 

\subsection{Background: singlet versus triplet}
The fully gapped superconducting state of LaNiGa$_2$, with critical fields\cite{Badger2022} H$_{c2}(0)$=0.27, 0.09, 0.24 T for uniform applied fields along the three ($a, b, c$) crystal axes (Appendix~\ref{app:SCparameters}), typical of low T$_c$ singlet SCs, must accommodate the `external' magnetic field from the muon via superconducting magnetic shielding currents, analogous to the simpler case of the Meissner effect at a surface. LaNiGa$_2$ is Type II,~\cite{Badger2022} -- anisotropic Abrikosov indices $\kappa$=3.38, 28.9, 4.00 along the three crystal axes -- like several other orthorhombic fragile  magnetic superconductors, listed in Sec.~\ref{subsec:TRSBlist}.

Standard Meissner expulsion of magnetic fields applies to singlet pairing, where only the supercurrent density is involved. Triplet OPs respond to magnetic fields very differently than do singlets, as shown from time-dependent Ginzburg-Landau simulations by Pechenik {\it et al.}\cite{Pechenik2002} They found that breaking of spin-rotation symmetry requires less energy than for singlet states, which implies a
paramagnetic contribution to susceptibility as opposed to the purely diamagnetic response of singlet pairing, which would enhance critical fields. Rosenstein {\it et al.} described that longitudinal and transverse (to the direction of the OP) penetration depths and coherence lengths can be  very different\cite{Rosenstein2015} for an isotropic 3D model, which would be significantly complicated for a crystal of orthorhombic symmetry. 
All materials parameters of LaNiGa$_2$, though crystal axis dependent, are in line with singlet pairing,\cite{Badger2022} with the exception of indications of additional depolarization from $\mu$SR studies (see Sec.~\ref{subsec:TRSBlist}). In a triplet SC, it can be expected that a magnetic field will (i) tend to rotate and align triplet pairs relative to the spontaneous SC magnetization, increasing magnetization into the applied field direction (increasing $|\uparrow\uparrow>$ pairs versus $|\downarrow\downarrow>$ pairs), engendering a possibly large spin susceptibility process not available to singlet pairing, and more general supercurrents may arise. Each of these effects, and more, may depend on the type of OP.\cite{Rosenstein2015}

To quantify the energy cost of converting spin-down pairs to spin-up pairs would require a specific Hamiltonian. Studies of quantified properties of triplet phases are at an early stage. One suggestion\cite{Rosenstein2015} is that coherence lengths can depend strongly on the choice of triplet OP. Unusual properties are not observed in LaNiGa$_2$.

\subsection{Non-symmorphic symmetry of LaNiGa$_2$}
\label{sec:FS}
Studies of single crystals of LaNiGa$_2$ revealed\cite{Staab2022,Badger2022,Quan2022} its space group to be $Cmcm$, with a non-symmorphic crystal symmetry, versus the 1982 assignment of symmorphic $Cmmm$ space group based on powder xray data.\cite{Grin1982} The structural similarity and differences
 have been provided in the original discussion.\cite{Quan2022} The non-symmorphic
space group operation results in a double band degeneracy (as well as spin symmetry)
across an entire face of the Brillouin zone, a well known non-symmorphic symmetry
consequence. A planar band degeneracy is an unusual occurrence in an exotic SC.
 with points on the Fermi surface with topological index thus characterizing it as an exotic topological superconductor.

It is useful to clarify the topological aspect. Upon including spin-orbit
coupling (SOC), this planar degeneracy is lifted throughout the zone including
on symmetry planes {\it except} along a single symmetry line, the $Z$-$T$
line).\cite{Badger2022,Quan2022}  Any pair of bands cutting the Fermi energy
along this line thus has a Dirac point degeneracy in the normal state at
that point. The Dirac point will lie {\it on the Fermi surface, independent
of doping or symmetry-preserving strain}, {\it i.e.} independent of the
position of E$_F$ -- $Cmcm$ symmetry demands the degeneracy along
the $Z$-$T$ line. The Dirac point simply moves along the $Z$-$T$ line,
only vanishing if the Fermi surfaces no longer cross the symmetry line.
It is highly unusual -- an accidental improbability -- to have {\it diabolical points
that remain pinned} to the Fermi surface as it varies due to external
influences -- doping, strain, pressure,etc. -- remaining on the Fermi surface
as long as the structural symmetry is retained. This symmetry-related pair of
points has four-fold degeneracy and the topological character of anisotropic
3D Dirac points. The Fermi surfaces, symmetry point labelings, and $Z$-$T$
line are shown in Fig.~\ref{fig:FS}.

This topological degeneracy is only lifted by the opening of the
SC gap.\cite{Badger2022}  Assuming a triplet OP, the
bands that are involved open with two bandgaps, given by the expression
(see Appendix~\ref{app:tripletOP}, Eq.~(\ref{eqn:BdG}) for the BdG quasiparticle bands).
This point degeneracy of a quartet of bands (including spin and SOC)
at a pair of Dirac points, finally
split only by pairing, form a basis with precise degeneracy for the discussion of a
two-component system in $k$ space. Previously, approximate symmetries
based on bands or Fermi surfaces have been suggested to allow for triplet pairing.
Ghosh {\it et al.} proposed Hund's coupling within Ni $3d$ orbitals to account
for triplet pairing,\cite{Ghosh2020} invoking symmetry-breaking of one of the
$3d$ orbitals. However, the density of states $N(E)$
indicates that the Ni $d$ bands are rather narrow and fully occupied,
lying in the -3 eV to -1.5 eV range and contributing only mildly at E$_F$
through mixing with Ga $s$-$p$ orbitals.\cite{Quan2022}


\subsection{LaNiGa$_2$ material parameters, energy scales}
\label{app:SCparameters}
The choice here is a representative case: the topological superconductor
LaNiGa$_2$ (orthorhombic, $Cmcm$), $a=4.29\AA, b=17.83\AA, c=4.27\AA$)
with the characteristic measured quantities given below taken
from the Supplemental Information of
Badger {\it et al.}\cite{Badger2022}, giving values for single crystal
samples. The analysis assumes conventional BCS singlet pairing.
This list is followed by several calculated or
estimated energies mostly from Quan {\it et al.}\cite{Quan2022} When three
numbers are given, they refer to the $a, b, c$ lattice direction
anisotropic values, and T=0 values are given for the T-dependent
properties. KWR is the Kadawaki-Woods ratio.\\
\vskip 2mm \noindent
- T$_c$=2 K;~~k$_B$T$_c$=0.17 meV, a relevant energy scale\\
- $\lambda_{GL}$= 174, 509, 189 nm \\
- $\xi_{GL}$    =~51.5,~~17.6,~~47.3 nm \\
- $\kappa$~~~   =~3.38,~28.9,~4.00~~~ \\
- $\gamma$=14.1  mJ/mole-K$^2$ (specific heat constant)\\
- $\Delta C(T_c)/\gamma T_c$=1.33; BCS value is 1.43\\
- $\rho_o$=5.2 $\mu\Omega$~cm, residual resistance; clean limit \\
- $H_{p}(0)$=3.66 T, the (extrapolated) Pauli limiting field \\
- $H_{c2}$=0.275,~0.094,~0.253 T, Helfand-Werthamer \\
- $H_c$=23 mT, thermodynamic critical field\\
- KWR: $A/\gamma_n^2$=1.28 ($[\mu\Omega$ cm/K$^2$]/[mJ/mol K$^2$]$^{2}$)\\
\vskip 2mm \noindent
The $a$-$c$ near isotropy of the superconducting parameters is evident, while 
the long $b$-axis values are quite different.
Sometimes useful is averaging over the orthorhombic directions (viz. for
making estimates or comparing with cubic materials):
$\xi_{GL}$$\sim$35 nm, $\lambda_{GL}$$\sim$300 nm, $\kappa$$\sim$4-29
(a moderately to strongly
Type II superconductor, depending on direction), H$_{c2}$$\sim$0.2 mT.
These values compare favorably with powder values\cite{Hillier2012}
$\lambda_{GL}^{ave}$=350 nm, $H_{c2}^{ave}$=0.4 T, $\xi_{GL}^{ave}$=38 nm. 

The following list gives material properties and 
representative energies for LaNiGa$_2$,
using conventional notation. The notation $\delta {\cal E}$ is the
energy difference (gain due to ordering) between normal and SC or magnetic
phases. \\
\vskip 2mm \noindent
- N(0)=3.25 states/eV-f.u.-both spins, the DFT value\\
- N$_{BCS}$(0)=1.63 states/eV per spin\\
- $\gamma_o$=7.66 mJ/mole-K$^2$\\
- el-ph $\lambda=\frac{\gamma}{\gamma_o}-1$ = 0.84\\
- $\Delta_o\approx$1.5 kT$_c$=0.25 meV;~~$2\Delta_o$=0.5 meV\\
- $\Delta {\cal E}_{SC}=\frac{1}{2}N_{BCS}(0)\Delta_o^2$=$5.1\times 10^{-8} eV$\\
- $B^{spon}$=0.2 G from $\mu$SR data\\
- $\mu_B B^{spon}$$\approx$8$\times$10$^-9$ eV [$\mu$SR]\\
- $m^{spon}$=0.012 $\mu_B$/f.u.=3$\times 10^{-3}$$\mu_B$/atom [$\mu$SR]\\
- $\frac{m}{N_{\uparrow}(0)}\approx$4 meV, exchange splitting of bands [$\mu$SR]\\
- $\Delta {\cal E}_{mag}$=$\frac{1}{4}$I$_{st}$$m^2$=6 $\mu$eV [$\mu$SR]
        (Stoner $I_{st}\approx$ 0.5 eV)\\

Note: the BCS paper uses the Fermi level DOS/spin, $N_{BCS}(0)$.
The last energy in the list is the cost, in conventional theory, of producing
a magnetization $m^{spon}$ in a non-magnetic material. The disparity 
in the various  energy scales is apparent, however the energy related to the
chosen spontaneous field $B^{spon}$ is comparable to the BCS 
condensation energy $\Delta {\cal E}_{SC}$.
The value of $\lambda$ from this theory/experiment heat capacity is somewhat larger
than expected for a 2K superconductor.


\subsection{The triplet order parameter; (non)unitary}
\label{app:tripletOP}
TRSB order parameters, {\it i.e.} emergence of magnetic polarizations, are
conventionally represented by a spin-triplet SC OP of the 
form\cite{Sigrist1991}
\bea
\Delta_k = \Delta_o~ \vec{d}_k\cdot\vec\sigma~(i\sigma_y),
\eea
where $\Delta_o(T)$ is the magnitude, $\vec{d}$ is a normalized complex triplet 
magnetization 3-vector whose $k$- and band $n$-dependence
is neglected until required by data, and $\vec\sigma$ is the vector
of Pauli matrices for spin. In terms of 
triplet spin notation this is\cite{Sigrist1991,Annett1995}
\bea
d_{\uparrow\uparrow}    &=&-d_x+id_y, \nonumber\\
d_{\downarrow\downarrow}&=&~~ d_x+id_y\nonumber \\
d_{\uparrow\downarrow}  &=&~~ d_z =d_{\downarrow\uparrow}.
\eea
(The vector $\vec d_k$ has nothing to do with a singlet 
$d$-wave order parameter symmetry.)

This straightforward extension to triplet order parameter, with spin-pairing
being even with respect to the Cooper pair that must be odd (antisymmetric) 
in total leaves
an additional odd symmetry order parameter to be included. 
The point group of LaNiGa$_2$ (also for LaNiC$_2$, with its
symmorphic but noncentrosymmetric space group) has only one-dimensional 
irreducible representations in the universe of possible symmetries
to be broken, hence providing no crystal symmetry
to be broken. This deficiency was addressed
by including an additional electronic degree of freedom, with the choice
being a similarity of two Fermi surfaces or a pair of bands. Neither of these
provides an exact underlying symmetry. The choice of a pair of atomic orbitals on 
(exactly) symmetry related Ni sites was also proposed. This proposition
confronts the observations that the Ni $3d$ bands are filled,
lying below -1 eV, with little opportunity for polarization, charge
order, or other types of symmetry breaking. 

Whatever the entities might be, the two-dimensional orbital/band
symmetry to be broken below T$_c$ is denoted by Pauli matrices
$\vec \tau$ in that space, for which some interaction might provide
quantum mechanical coupling. 
The INT order parameter\cite{Weng2016,Ghosh2022} becomes
\bea
\Delta_k = \Delta_o~ \vec{d}_k\cdot\vec\sigma~ (i\sigma_y) \otimes (i\tau_y),
\label{eqn:fulltripletOP}
\eea 
where $\vec\tau$ incorporates the two level band, Fermi
surface, or orbital (near)degeneracy. 
This form is the one currently proposed for LaNiGa$_2$.

Once included in the BdG quasiparticle Hamiltonian,
the eigenvalue dispersions  $E_k$ are\cite{Sigrist1991}
(incorporating the magnitude $\Delta_o$ into the $\vec d$ vector)
\bea
E_k&=&\pm\sqrt{\varepsilon_k^2  +\Delta_o^2(\vec d\cdot\vec d^* \pm
      |\vec d\times\vec d^*|)} \nonumber \\
   &=&\pm\sqrt{\sum_{j=1}^3 v_{k,j}^2 \delta k_j^2 
    +\Delta_o^2(1 \pm |\vec d\times\vec d^*|)`},
\label{eqn:BdG}
\eea
where $\varepsilon_k$ is the normal state band energy expanded 
in $\delta k_{j}$ around the DP.
The cross product is imaginary or zero.

\begin{figure}
 \includegraphics[width=1.0\linewidth]{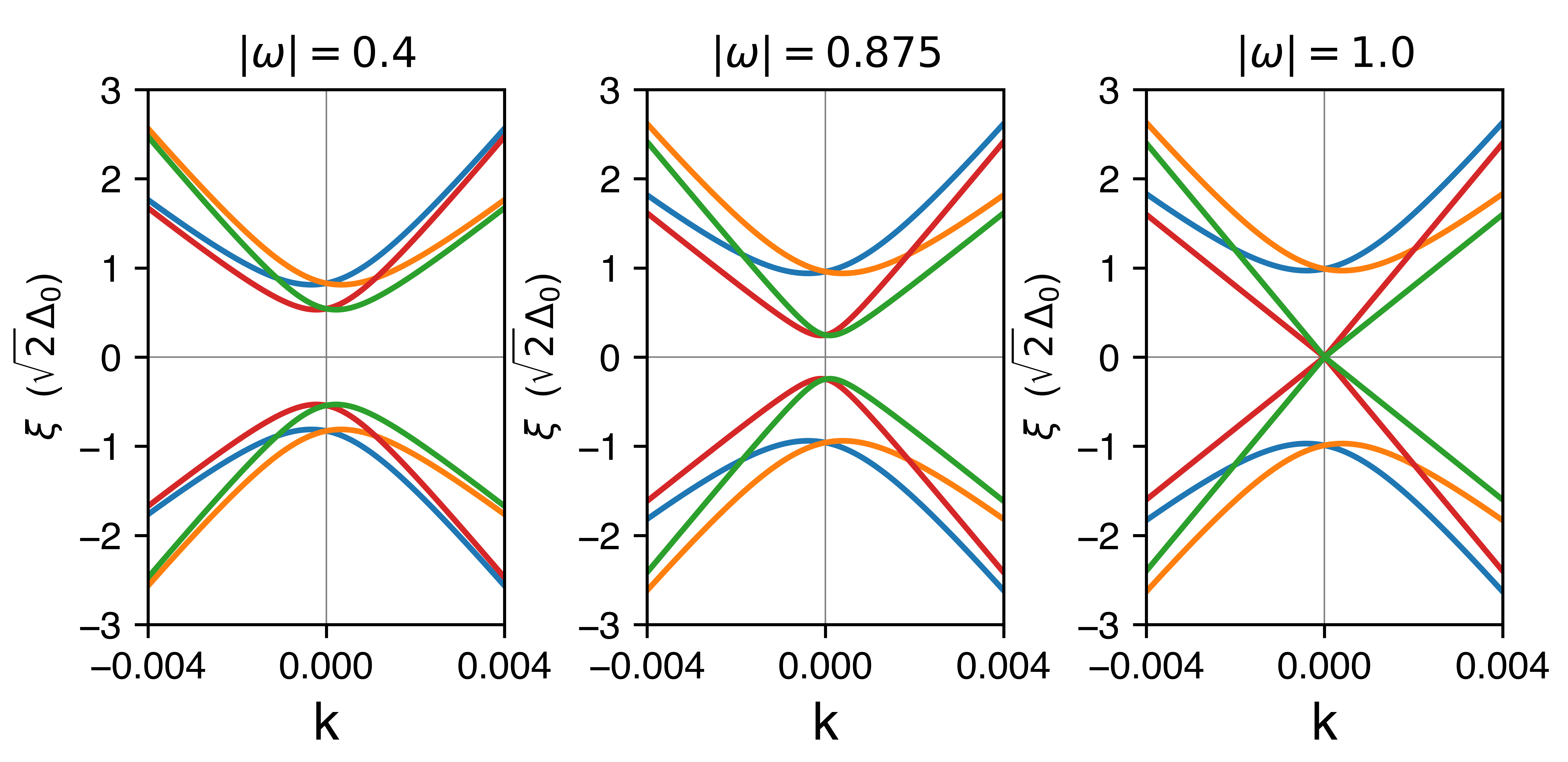}
\caption{Sketch of the quasiparticle  (Bogoliubov-de Gennes)
 band structure along one dispersive
band direction of the triplet, topological superconductivity
model for LaNiGa$_2$.\cite{Quan2022} Graphs for three values of the 
``triplet strength'' $\omega$ are displayed. For $\omega$=1,
the non-symmorphic band sticking at the zone boundary, hence two
nodes, persists into the then gapless superconducting state. In the nonunitary
regime $\omega<1$ the degeneracy is broken, giving a gap to
quasiparticle excitation.
}
\label{fig:QP-dispersion}
\end{figure}

{\it (Non)unitarity more generally.}
Study of exotic OPs introduces the property of unitarity, or not.
From Eq.~\ref{eqn:BdG} an obvious special case is when the cross
product is unity, for which two of the eigenvalues are not split --
a degeneracy persists. An instructive example is
$\vec d=\Delta_o (\cos\eta,i\sin\eta,0)$ for the usual range of 
phases $\eta$. The cross product is (0,0,-i sin2$\eta$), which is
unity (``unitary'') if $\eta$=$\pi/4$ or $3\pi/4$, otherwise it 
is less than unity (``non-unitary'')
except for the trivial cases of $\eta$=0 or $\pi/2$. The difference 
is the relative phase difference between components of $\vec d$,
which may be a delicate physical property. 
Figure~\ref{fig:QP-dispersion} presents a sketch of the quasiparticle
bands around the Dirac point of LaNiGa$_2$ for the unitary case
and for two magnitudes of non-unitarity 
  $\omega=|\vec d \times \vec d^*|$, only the unitary case
is gapless. Experimentally, LaNiGa$_2$ is fully gapped, {\it i.e.}
non-unitary, if this form of OP holds.

In the formulation of Ramires 
and coauthors, their time-reversal operator is defined by
\bea
\vec q_{tro} = 2i~\vec d_k\times\vec d_k^*.
\eea
If nonunitary,  TRS has been broken. Additional
information, related to spin-orbit coupling (with or without), 
and effects of distinct  point
groups, are available from Ramires' paper.~\cite{Ramires2022}
The symmetry classification and properties of triplet OPs, generalized
from the spherically symmetric case of $^3He$ to crystal symmetry,
 was presented by Blount in 1985.~\cite{Blount1985} 
For nonunitary consequences of crystal
symmetry, an overview appeared by Ramires,\cite{Ramires2022} after extensions
of the theory and some proposed examples from experiment.  
Sigrist and Euda\cite{Sigrist1991} expressed the criterion 
for unitarity of a general OP as 
\bea
\Delta_k \Delta_k^{\dag} \propto \sigma_o.
\eea
That is, if the generalized magnitude of $\Delta_k$ is proportional to
the identity matrix, the state is a unitary one. Here this means that
 if $i\vec d^*_k \times \vec d_k$ is non-zero ({\it i.e.}
they are `non-parallel' in their complex vector space), then the OP is
non-unitary. This property 
results in a non-vanishing average of the magnetic moment
$\Delta_k^{\dag}\vec\sigma \Delta_k$.\cite{Sigrist1991} A related OP
couples to an external magnetic field in the Ginzburg-Landau
free energy functional.\cite{Bao2013}

\subsubsection{Comments about magnetic superconductors}
A charged fermionic superfluid, due to orbital effects, is conspicuous in
its insistence on expelling a magnetic field (Type I), or confining it to
vortices (Type II). In the singlet case the SC order parameter 
is depressed and vanishes at the vortex center, for triplet the behavior
can become modified.\cite{Rosenstein2015} 
For triplet SCs there is the issue of their own self-magnetic
field due to spin imbalance, which
presumably survives (remains frozen in) in the steady state 
description of triplet SC properties. The behavior of a
triplet SC in its own field, even more so in an applied field, seems to be little studied
theoretically -- apparently depending on the specific type of triplet
OP -- and not yet resolved by experiments.

This puzzle is most prominent in strongly ferromagnetic (FM) SCs such
as UGe$_2$, where FM magnetic order results from large moments
(up to 0.9$\mu_B$) on the U atoms, whose
magnetic origin is evidently  of conventional
electronic exchange interactions and whose ground state electronic
structure is described reasonably
(but probably not conclusively) in the bulk by
correlated DFT methods, see for example 
Refs.~[\onlinecite{Shick2004},\onlinecite{Shick2021}].  The appropriate
description of $5f$ electrons (localized versus itinerant, treatment of
spin-orbit coupling [$\vec L\cdot\vec S$ versus $\vec J\cdot\vec J$], or more
detailed correlations) seems to depend on band filling ({\it i. e..} the 
progression along the row of actinides), pressure, doping, and several other
material characteristics. Orbital moment scenarios are mentioned
briefly in Sec.~\ref{sec:orbitalmoment} but are not appropriate for
large magnetic moment SC states.

Inhomogeneities complicate this already uncertain picture.
Relevant inhomogeneities include surfaces and interfaces, domain boundaries,
and planar, line, and -- in particular -- point defects,
with interstitial muons being the most studied type of point magnetic defect. 
Magnetic regions in nominally nonmagnetic actinide-based materials
have been discussed as a material complication. These disruptions likely lead to local
supercurrents that produce a magnetic field in the region, in a magnetically
susceptible metal. Local magnetic regions have been suggested as the origin
of polar Kerr rotation observed in U-based SCs.
TRSB fields, when observed, by $\mu$SR, by polar Kerr effect, or by
magnetization, are not far from the threshold of detectability. 
Magnetization in LaNiC$_2$,\cite{Sumiyama2015} reported as resulting from a
field of 0.01 G, is the most noticeable case.

The Meissner effect might have some theoretical commonality
for singlet and triplet pairing, as it involves only an orbital (charge) component.
The formulation of a microscopic theory of orbital currents, which
possess a local angular momentum and provide a local moment, 
in a periodic solid has not been a straightforward theoretical question.
Robbins {\it et al.}~\cite{Robbins2020} have applied developments in the
theory of polarization in crystals to address this challenge for the
normal state, then extended the theory of the
orbital moment in a superconductor to the formulation and application of
a modern DFT-based band structure code (viz. BdG formulation for a SC) to
explore such effects in real materials.

This work of Robbins {\it et al.} required a modern
formulation of orbital momentum because in the usual angular momentum
operator $\vec L = \vec r \times \vec p$, the position operator $\vec r$
is tricky to handle in an extended, periodic system, while the orbital
moment will be a local or periodic quantity. Assuming a chiral, $p$-wave OP
for the correlation-enhanced and enigmatic compound Sr$_2$RuO$_4$, the model of
Robbins {\it et al.} led to an predicted orbital moment
of 3$\times 10^{-4}\mu_B$
per f.u., which they suggest accounts for a field of 0.3 G, 
near to the detection limit of zero-field $\mu$SR studies.
The formalism they develop can be expected to be important in modeling
and understanding orbital moments in superconductors.
This topic of orbital currents is discussed
in Sec.~\ref{sec:orbitalmoment}, including the specific proposal
by Ghosh {\it et al.}\cite{Ghosh2021d} of TRSB loop supercurrent order.



\section{Supplementary Material:
Time Reversal Symmetry Breaking and
{\it Fragile Magnetic Superconductors}}
\noindent
Warren E. Pickett \\
Department of Physics \& Astronomy \\
University of California Davis, Davis CA 95616\\
\vskip 4mm

\noindent
{\bf Reference numbers correspond to the Reference List in the main text.}
\date{\today}
\maketitle

\section{{The fundamental Hamiltonian}}
\label{SM:hamiltonian}
Since supercurrents are determined by a vector potential, it is not
sufficient to consider the Hamiltonian only at the magnetic field level.
With the muon producing a vector potential $\vec A^{\mu}$ in an electron gas,
and allowing for a spontaneous vector field $\vec A^{spon}$ appearing below T$_c$,
the non-relativistic Hamiltonian for the muon and the conduction electrons is
(leaving the subscript $e$ off electron operators)
\begin{eqnarray}
H^{orb}&=& \frac {[p_{\mu}+\frac{e}{c} \vec A^{spon}(\vec r_{\mu})]^2}{2m_{\mu}}
          +\sum_{j,s} \frac {[p_{j,s}-\frac{e}{c} \vec A(\vec r_{j,s})]^2}{2m} \nonumber \\
       & &  -\sum_{j,s} \frac {e^2} {|\vec{r}_{j,s}-\vec{r}_{\mu}|}
             +V^{e-ion} + V^{e-e} \nonumber \\
       & &   +H^{\mu-ion} + T^{ion} + V^{ion-ion} \nonumber \\
\vec A(\vec r)&=&\vec A^{\mu}(\vec r) + \vec A^{spon}(\vec r)\nonumber \\
  \vec A^{\mu}(\vec r)&=&\nabla\times\frac{\vec\mu}{r}
                       =\frac{\vec\mu\times \vec r}{r^3}.
\end{eqnarray}
Several terms are abbreviated with self-evident notation.   
In an applied field the corresponding vector potential $\vec{A}^{ext}$
would be included in each vector potential term. The electronic terms are
to be treated with DFT methods, and the ion kinetic energy $T^{ion}$
vanishes for static atoms (and is treated classically in determining the
relaxed structure). Treating the vector potential accurately in density
functional terms requires
current density functional theory,\cite{Vignale1990} which has not been implemented
in current codes.  Nuclear vector potentials are not displayed here, nor
are those of the itinerant electrons which without polarization are
presumed to average out, randomly or statistically, so the precise
Hamiltonian has a more involved form that is rarely addressed.
Each vector potential field is evaluated at the position of the respective moment.
The first term is for the muon,
the second is the kinetic energy for all electrons. The muon does not
experience its own field.
The curl operators provide the magnetic fields that couple to the charges and moments.

Moving to Dirac's relativistic treatment of the kinetic energy operator
results in expressions that involve the particle (muon and
electrons) spin moments [again, nuclear moments are neglected here although
they contribute to the various fields that are experienced by the muon) 
that couple to magnetic fields (other than their own),
external or from other particles.
The spin-related terms are
\bea
H^{spin}&=& -\mu_B \sum_{j}\vec\sigma_j\cdot\nabla\times
    (\vec A^{\mu}+\vec A^{ind}+\vec A^{spon}) \nonumber \\
  &  &+~~\vec \mu \cdot \nabla\times(\vec A^{ind}+\vec A^{spon}),
\eea
where $\vec\sigma_j$ is the Pauli spin matrix of the $j$-th electron.

The quadratic term (in $\vec A^{\mu}$) in the kinetic energy operator has the form
\bea
H^{(2)}&=&\frac{e^2}{2mc^2}  \vec A^2=\frac{\mu_B^2\mu^2}{\hbar^2}\frac{(x^2+y^2)}{r^6}\nonumber \\
       &=&\frac{\mu_B\mu^2}{\hbar^2}  [\frac{\rho^2}{(\rho^2+z^2)^3}].
\eea
The last expression is written in the natural cylindrical coordinates
$\rho$ and $z$, with absence of the
polar angle $\phi$ reflecting the circular symmetry.
This term falls off as $r^{-4}$; on the other hand, it diverges with a
power as $r$ approaches the muon position, becoming worse as $z$
approaches zero; there is however an extra factor of $\mu$.
Final effects are apparently negligible and are not addressed in this paper.

\section{The muon magnetic operator}
\label{SM:muonmoment}
The linear-in-field term involving the muon in the Hamiltonian simplifies readily.
For each electron $j,s$, and dropping this subscript for simplicity,
($\vec A^{\mu}$ is evaluated at $\vec r_{j,s}$; $\vec p=-i\hbar\nabla$)
\begin{eqnarray}
H^{\mu}_B&=&-\frac{e/c}{2m}
   \big[\vec p\cdot\vec A_{\mu}+\vec A_{\mu}\cdot\vec p\big]\nonumber \\
  &=&  \frac{2\mu_B}{\hbar} \vec A_{\mu}\cdot\vec p,
\end{eqnarray}
using the definition of the Bohr magneton $\mu_B=e\hbar/(2mc)$ and noting that in the
divergenceless gauge $[\vec p\cdot \vec A^{\mu}]$
vanishes identically.
Then the magnetic part  of the Hamiltonian becomes (for each electron)
\begin{eqnarray}
H^{\mu}_B&=& \frac{2\mu_B}{\hbar} \vec A^{\mu} \cdot \vec p \nonumber \\
         &=& \frac{2\mu_B}{\hbar}(-i\hbar) \frac{\mu}{r_3}(-y,x,0)\cdot
                (\partial_x,\partial_y,\partial_z)          \nonumber \\
         &=&-\frac{i\mu_B\mu}{r^3}(-y\partial_x + x\partial_y)\nonumber \\
         &=&  \frac{\mu}{r^3}\mu_B \frac{L_z}{\hbar} \rightarrow
             \frac{\mu\mu_B}{\hbar}<\frac{L_z}{r^3}>.
\end{eqnarray}
This expression contains the $\mu/r^3$ factor of $\vec B^{\mu}$~
(Sec.~\ref{sec:near-field}) leading to a troublesome radial integral
to quantify the perturbation.
$\vec L$ is the electronic orbital angular momentum operator relative
to the muon at the origin. The Hamiltonian thus favors the development
of an electronic angular momentum with respect to the muon position.
This may produce an orbital magnetic moment, allowed by the broken
TRS due to the muon magnetic moment.

The vector potential accompanying the $\mu^+$ magnetic moment,
which does not affect the muon that creates it, is conventionally
treated, again in the Coulomb gauge $\vec{\nabla}\cdot \vec A=0$, as
\begin{eqnarray}
 \vec A_{\mu}(\vec r)=\nabla\times\frac{\vec\mu}{r}=
    \frac{ \vec {\mu}\times \hat{r}}{r^2}  
 =\frac{\mu}{r^3}(-y,x,0),
\end{eqnarray}
a toroidal potential field with a divergence that displays a
singularity at the muon site. One first considers the ground
state: the muon sits in a stable quadratic potential in its harmonic
oscillator ground state with its spin along the $z$-axis, and the
electrons are in, or very near, their ground state.

The dipole magnetic field, from above, is (neglecting the Fermi
contact $\delta$-function term\cite{Autschbach2012})
\begin{eqnarray}
\vec{B}_{\mu}(\vec r)&=&\nabla\times\vec{A}(\vec r) \nonumber \\
  &=& \frac{ 3\hat{r} (\hat{r} \cdot \vec{\mu})-\vec{\mu}}{r^3}
              \nonumber \\
  &=& {\mu} \frac{ 3\hat{r}(z/r) -\hat{z} }{r^3}   \nonumber \\
  &=&\frac{{\mu}}{r^3}
     \big(3\frac{xz}{r^2}, 3\frac{yz}{r^2}, 3\frac{z^2}{r^2}-1\big).
\end{eqnarray}

Treated as a perturbation, the first order change in energy of an
electron in state $\eta_{n\ell m_{\ell}}(\vec r)$ in a spherical potential is
\bea
\Delta E&=&\big<\eta_{n\ell m_{\ell}}|
 \frac{\mu}{r^3}\mu_B \frac{L_z}{\hbar}|\eta_{n\ell m_{\ell}}\big>\nonumber \\
   &=&\mu\mu_B m_{\ell} \big<n\ell m_{\ell}|\frac{1}{r^3}|n\ell m_{\ell}\big>
        <\ell m|L_z|\ell m>.
\eea
This expression as written is indeterminate for $s$ orbitals: the orbital is finite
at the origin so the integral is $\int r^2 dr/r^3$, thus logarithmically divergent
if the $dr$ integral is done first. The $m_{\ell}=0$ for $s$
orbitals would give a zero result if the angular integral were to be
done first and the environment has circular symmetry.
For higher $\ell$ states $p, d, f, ...$, the wavefunction is
proportional to $r^{\ell}$, thus giving an extra $r^{2\ell}$ factor
in the integrand near the origin and a finite integral.
Since hyperfine quantities are the topic here, one can note that a similar
question arises in a relativistic treatment (Dirac equation), where the
$p_{1/2}$ orbital is non-zero at the origin.

A successful treatment of this indeterminate integral was given by Abragam and
Bleaney.~\cite{Abragam2012} Reverting to the relativistic Dirac equation,
the integrand for the matrix element (evaluated with the large component
of the wavefunction $\gamma(\vec r)$) was manipulated using the expression
$\vec A=\nabla\times(\vec\mu/r)$. The perturbation term is
\bea
\frac{\beta}{\hbar}\vec\sigma\cdot\big<\gamma|[\vec A\times\vec p +
     \vec p\times\vec A]|\gamma\big>
\eea
where $\beta$ is the 4$\times$4 Dirac $\beta$-matrix. The gradient operator
operates on the function $\gamma$, and $\gamma^*\nabla\gamma$ is
one half of the density gradient $\nabla \gamma^*\gamma$. Thus the
derivative operator that had non-relativistically resulted in a
$r^{-3}$ factor in the denominator has been transferred to the gradient
of the (large component) spin current
\bea
\vec J_{spin}= \nabla \times |\gamma(\vec r)|^2\vec \sigma,
\eea
leaving only a $r^{-2}$ factor in the denominator of the integrand,
which is regularized
by the $r^2 dr$ volume factor. A key factor is the
replacement the usual expression of the vector potential by the
differential form $\vec A = \nabla\times\frac{\vec \mu}{r}$.
The ground state orbital (above denoted by $\gamma$) is asymmetric
for a general non-symmetric position of the muon, and will include
$p, d, ...$ symmetry contributions in a spherical expansion. The
$s_{1/2}$-$p_{1/2}$ contribution to the gradient of the spin density
will be regularized in the same manner as the diagonal $s$ treated just
above, and higher-$\ell$ contributions will approach zero in a way
that the integrals are straightforward and finite.

\section{$\mu$SR: experiment and analysis}
\label{SM:muSR}
\subsection{The setup: {$\mu^+$} + sample}
Overviews and reviews of the experiment and analysis were referenced in
Sec.~\ref{sec:intro}. A synopsis is given here.
The $\mu^+$ ion, an elementary particle, decays after production
with a half-life of $\tau_{\mu}$=2.2 $\mu$s, producing a positron (conserving
charge) and two neutrinos (one anti-electron type, one muon type, conserving
lepton number); the entire process conserves the relativistic 
4-momentum). Positron emission occurs symmetrically around the direction of
the muon spin, but with the variation in angular distribution changing with
the energy of the positron. The mass of the
muon provides $\sim$100 MeV of energy for the decay products, of which the
positron carries away 25-50 MeV kinetic energy (its rest mass
of 0.5 MeV can be neglected). At the lower end of
emission energy, ~25 MeV, emission is practically isotropic -- at any angle
with respect to the muon spin. Toward the higher range of energy $\sim$50 MeV,
the angular dispersion of emission becomes weighted toward the
forward direction (the direction of the $\mu^+$ spin at the time of decay),
although with little variation within a cone of $\pm$40$^{\circ}$ of forward.
These two distributions are pictured in Fig.~\ref{fig:PositronEmission}.

The spin-half moment has the largest quantum uncertainty of its direction
(a mathematical concept, if not a physical one, since only $s_z$ is
specified). Thus for full polarization $s_z=+1/2$, $\vec s$ varies off-vertical
by $\sin^{-1}(s_z/|\vec s|)=\sin^{-1}(1/\sqrt{3})$$\approx$35$^{\circ}$ with random polar
angle, a variation that must contribute to the distribution that varies
little within the $\sim$$\pm$$40^{\circ}$ forward cone and increases as
depolarization occurs. It is difficult to find this quantum uncertainty of
spin direction addressed in the descriptions of analysis.

The direction of emission of the positron, described
as predominantly forward (muon $s_z$ direction)
at the time of decay, is useful to understand. Decay of the positron
along the muon spin direction must take into account quantum spin
``direction'' mentioned above, and that the emission direction is
strongly dependent on the positron kinetic energy $E_p$.
Defining the energy parameter $\zeta=E_p/m_{\mu}c^2$, denoting
the fractional polarization at time of decay as $P_{\mu}$, and using
$\theta$ as the angle between spin at decay relative to the initial
polarization, the decay distribution from the quantum field theory of
particles according to Bayes is\cite{Bayes2011}
\bea
\frac{d^2\Gamma}{d\zeta d\cos\theta} &\propto& \zeta^2\Bigl[(3-3\zeta)+\frac{2}{3}
  \rho(4\zeta-3)+3\eta\zeta_o\frac{1-\zeta)}{\zeta} \nonumber \\
 & & +P_{\mu}\xi\cos\theta[(1-\zeta)+\frac{2}{3}\delta(4\zeta-3)]\bigr].
\eea
The constants are Standard Model parameters, calculated by
Michel\cite{Michel1,Michel2}
to be  $\rho=\delta=\frac{3}{4}, \xi=1, \eta=0$.
There is no reason here to try to fully comprehend
this distribution, the point is that it is a necessary Standard Model result and there has
not yet been any known violation of the SM. These calculated values have been
verified experimentally\cite{Grossheim2008} within very tight uncertainties,
and considerably simplify the expression.

\begin{figure}
 \includegraphics[width=0.6\linewidth]{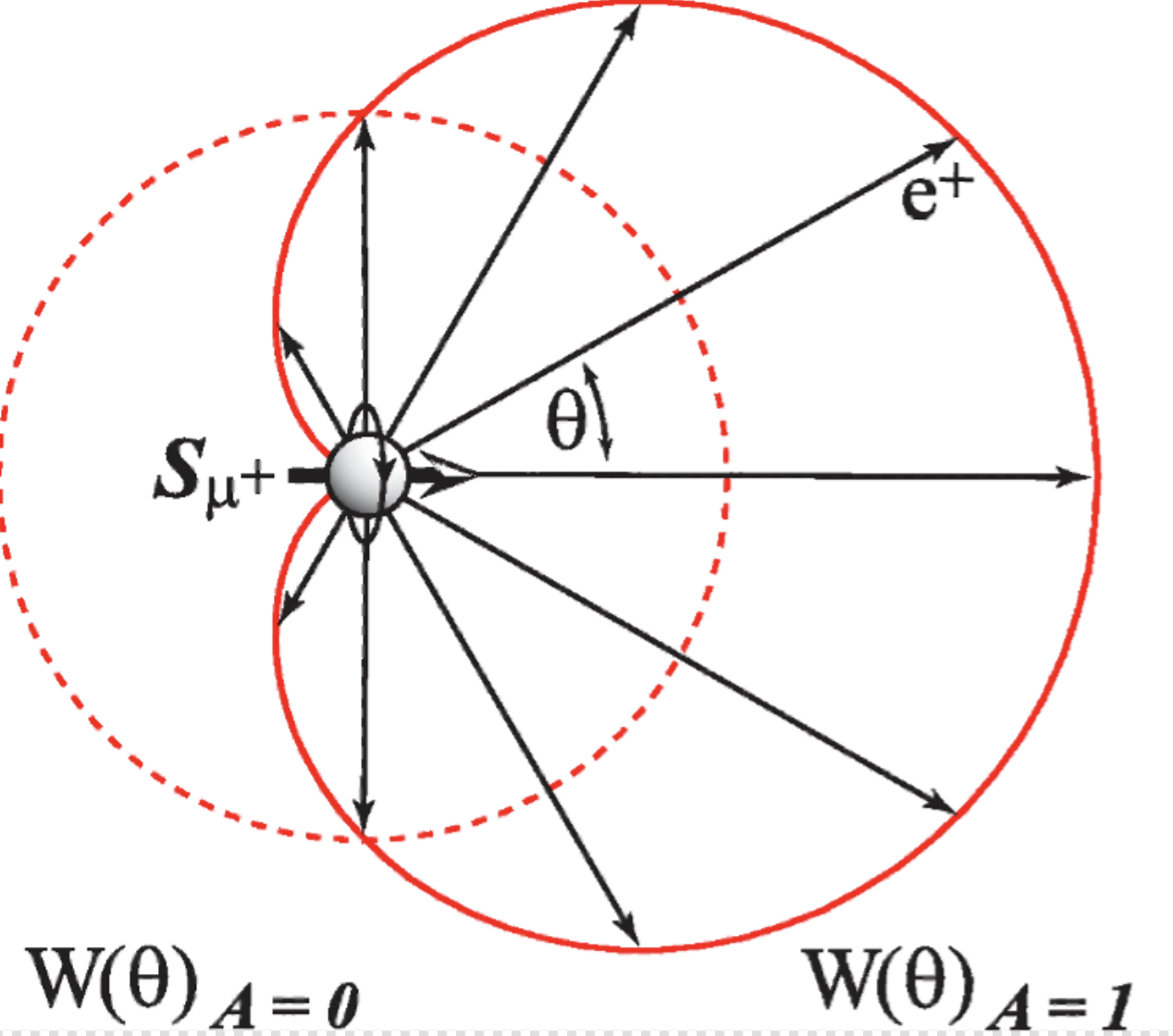}
\caption{Repeated from the main text for handy reference.
Angular distribution of the direction of emission of the
positron upon muon decay into a positron and two neutrinos, given
by $W(\theta)\propto 1+A\cos\theta$. The factor $A$ depends on the
energy of the emitted positron.
Two positron energies are pictured: dashed line, 25 MeV near minimum,
$A\approx 0$; full red line, 50 MeV near maximum, $A\approx 1$.
The arrow denotes the muon  polarization
direction upon decay. For reference: the rest mass of the muon
is $207\times 0.511$ MeV $\sim$ 100 MeV, which is what is
available for the decay products. See text for further discussion.
}
\end{figure}

Measurements of counts and angles
by an array of detectors
enables collection of counts versus time that contains
the information on depolarization.
Typical solid state processes are orders of magnitude quicker than the
$\mu^+$ lifetime, thus allowing accumulation of data representative of the
muon within an electronic system in its thermodynamic state at
temperature T and time $t$ after deposition.

\subsection{$\mu$SR: data \& analysis}
\label{subsec:analysis}

The direction of the muon polarization defines our $z$ direction.
The muon's spin is associated with a point magnetic dipole $\mu$
at the muon site, in a favorable interstitial stopping site,
away from positive atomic cores, and in metals not pictured
as being bonded with an atom.

The muon is like the much studied proton interstitial,
but with smaller mass and larger moment.
Takada, Maezono, Yoshizawa\cite{Takada2015} have studied, with density
functional methods supplemented by diffusion Monte Carlo calculations,
the case of a proton in an electron gas from high density, where the
system becomes that of a proton mixing with and screened by electrons,
to the low density
regime where the proton attracts an extra electron to become an H$^-$
ion favored by the filling of the $1s$ shell. The intermediate regime
includes the Kondo effect. Takada {\it et al.} did not consider
explicitly the vector potential of the proton.
The discussion in this paper is confined to intermetallic metals with typical
metallic densities; semimetals and insulators require their own treatment.

Typically presented data from zero-field studies is the rate of depolarization
$\sigma(T)$. Above $T_c$ there is a normal state temperature
independent depolarization
from small quasi-static nuclear moment fields and from larger
fluctuating magnetic fields at a continuum of distances and with
random directions of moments. These are
observed to average to some
constant rate above T$_c$.  Below $T_c$ in
materials identified with TRSB, the rate increases with decreasing temperature
to a value as $T\rightarrow 0$ that is
$\sim$3-10\% above the normal  state value.  $\sigma(T=0)-\sigma(T)$
is interpreted in terms of depolarization arising from an emergent field
felt by the muon, with the simplest interpretation being that it arises
from a spontaneously appearing uniform magnetic field with direction
non collinear with the muon spin..

The key result in zero-field experiments is the time development
(relaxation) of the muon moment polarization, followed by the process of
muon decay. Distinctions may be required for the two types of muon
beams: continuous wave (CW), at
the Paul Scherrer Institute (Switzerland) and TRIUMF (Canada), and
pulsed, at ISIS (U.K.) and KEK (Japan),\cite{Blundell1999}
but these distinctions will not be considered here.

Theories of the depolarization process
have been presented at a few levels over time, of which three are mentioned
here. Very often the 1967 Kubo-Toyabe functional
form,\cite{Kubo1967} which takes into account a stochastic depolarization
field and anisotropy, is adopted as a first step in the process to
fit the depolarization data and extract materials
properties. The Kubo-Toyabe form gives the polarization distribution
$G^{KT}$ versus time as
\bea
A(t)       &=& A_0 G^{KT}_z(t) e^{\lambda_{ZF}t}, \nonumber \\
G^{KT}_z(t)&=& \frac{1}{3} + \frac{2}{3} (-\sigma^2t^2) e^{-\sigma^2t^2/2},
\eea
where $\sigma=\gamma\Delta$ in terms of an isotropic Gaussian
distribution width $\Delta$ and the muon gyromagnetic ratio $\gamma=2\pi
\times 13.55$ kHz/G. This form is used to fit data in many $\mu$SR papers. 
The asymmetry data is provided by
\bea
A^{expt}(t)=\frac{N_F(t) - \alpha N_B(t)}{N_F(t)+\alpha N_B(t)},
\eea 
where $N_F$ and $N_B$ are the time histograms of the temporal dependence
of the decay positron count rate in the forward and backward detectors
and $\alpha$ is an apparatus dependent calibration constant.  Schematic
illustrations of $F(t)$ and $B(t)$ were provided by Blundell.\cite{Blundell1999}

Another expression for analysis was included in the Topical Review of
Ghosh {\it et al.}\cite{Ghosh2021} The polarization function is given
by a generalized Lorentzian and Gaussian
Kubo-Toyabe form used for fit to data  is
\bea
A(t))=\frac{1}{3} A_0 + \frac{2}{3} (1-\sigma_{ZF}^2t^2-\lambda_{ZF})
         e^{-\frac{\sigma^2t^2}{2} -\lambda_{ZF}t}
\eea
where $A_0$ is the initial asymmetry and ZF indicates a zero field value.

The cases of LaNiGa$_2$ (T$_c$=2K) and LaNiC$_2$ (T$_c$=2.7K) provide
interesting insight into analysis. According to Ghosh {\it et al.}\cite{Ghosh2021},
evidence for TRSB occurs due to increase in $\sigma$ below T$_c$,
but sometimes appears as a change of $\lambda_{ZF}$ below T$_c$. For
LaNiGa$_2$ the change occurred in $\sigma$ whereas for LaNiC$_2$ the
change appeared in $\lambda_{ZF}$. Both have been included in the list
of TRSB superconductors. TRSB was also indicated by
direct magnetization measurement on single crystal LaNiC$_2$, where
a field was detected but only at the 0.01 mG
(10$^{-9}$T) level, and oriented along the $c$ axis.\cite{Sumiyama2015}
 The field onset occurred {\it above}
T$_c$=2.7K, in the 2.7-3.0K temperature region.

Kornilov and Pomjakushin\cite{Kornilov1991} in 1991 extended the theory to include
randomly-directed field with $\delta$-function strength
$H_o$ (possibly modeling a polycrystalline sample)  as well as the
Gaussian distribution. The analytic expression they obtained was
\bea
G^{KP}_z(t)&=&\frac{1}{3} + \frac{2}{3}\big[1+\frac{\sigma^2t}{\omega}^2\big]
                  e^{-\sigma^2t^2/2}\nonumber \\
         &\times& \big[\cos\omega t + \tan^{-1}\frac{\sigma^2t}{\omega}\big].
\eea 
This formula introduces a frequency $\omega=\gamma H_o$, and
plots of the difference from $G^{KT}_z$ versus $\sigma t$ were presented.

A recent (2020) description
has been given by Takahashi and Tanimura\cite{Takahashi2020} that is
said to consist of a more fully quantum derivation, and includes the thermal bath of
vibrations and spins, hence giving a T-dependent expression resulting
from their hierarchical equation of motion method. The result
requires numerical solution and fitting of several parameters, hence
the identification of a specific field value from experimental data is less
direct. 

{\it An aside from the Standard Model.}
The direction of emission of the positron, described
as primarily forward along
the muon polarization direction
at the time of decay, is more interesting, being
dependent on the positron kinetic energy $E_p$.
Defining the energy parameter $\zeta=E_p/m_{\mu}c^2$, denoting
the fractional polarization at time of decay as $P_{\mu}$, and using
$\theta$ as the angle between spin at decay relative to the initial
polarization, Bayes provided the decay distribution
from the quantum field theory of particles as\cite{Bayes2011}
\bea
\frac{d^2\Gamma}{d\zeta d\cos\theta} &\propto& \zeta^2\Bigl[(3-3\zeta)+\frac{2}{3}
  \rho(4\zeta-3)+3\eta\zeta_o\frac{1-\zeta}{\zeta} \nonumber \\
 & & +P_{\mu}\xi\cos\theta[(1-\zeta)+\frac{2}{3}\delta(4\zeta-3)]\bigr].
\eea
The constants are Standard Model parameters, calculated by Michel\cite{Michel1,Michel2}
to be  $\rho=\delta=\frac{3}{4}, \xi=1, \eta=0$.
There is no reason here to try to understand
this distribution, the point is that it is a necessary Standard Model result and there has
not yet been any known violation of the SM. These calculated values have been
verified experimentally,\cite{Grossheim2008} and considerably simplify
the expression.

\end{document}